\documentclass[%
aip,
jap,%
amsmath,amssymb,
reprint,%
]{revtex4-1}

\usepackage{graphicx}
\usepackage{dcolumn}
\usepackage{bm}

\usepackage{mathrsfs}
\usepackage[utf8]{inputenc}
\usepackage{amsmath}
\usepackage{amsfonts}
\usepackage{amssymb}
\usepackage{braket}
\usepackage{subfig}
\usepackage{color,soul}
\usepackage[colorlinks=true,linkcolor=blue,urlcolor=blue,citecolor=blue,breaklinks=true]{hyperref}
\usepackage[outercaption]{sidecap}

\newcommand*{\citen}[1]{%
	\begingroup
	\romannumeral-`\x 
	\setcitestyle{numbers}%
	\cite{#1}%
	\endgroup   
}

\usepackage{soul}
\usepackage{subfig}

\begin{document}

\preprint{AIP/123-QED}

\title{Optoelectronics of Inverted Type-I CdS/CdSe Core/Crown Quantum Ring}

\author{Sumanta Bose}
\email{sumanta001@e.ntu.edu.sg}
\affiliation{School of Electrical and Electronic Engineering, Nanyang Technological University, 50 Nanyang Avenue, Singapore 639798, Singapore; OPTIMUS, Centre for OptoElectronics and Biophotonics, Nanyang Technological University, 50 Nanyang Avenue, Singapore 639798, Singapore and LUMINOUS! Centre of Excellence for Semiconductor Lighting \& Displays, Nanyang Technological University, 50 Nanyang Avenue, Singapore 639798, Singapore}

\author{Weijun Fan}
\email{ewjfan@ntu.edu.sg}
\affiliation{School of Electrical and Electronic Engineering, Nanyang Technological University, 50 Nanyang Avenue, Singapore 639798, Singapore; OPTIMUS, Centre for OptoElectronics and Biophotonics, Nanyang Technological University, 50 Nanyang Avenue, Singapore 639798, Singapore and LUMINOUS! Centre of Excellence for Semiconductor Lighting \& Displays, Nanyang Technological University, 50 Nanyang Avenue, Singapore 639798, Singapore}

\author{Dao Hua Zhang}
\email{edhzhang@ntu.edu.sg}
\affiliation{School of Electrical and Electronic Engineering, Nanyang Technological University, 50 Nanyang Avenue, Singapore 639798, Singapore; OPTIMUS, Centre for OptoElectronics and Biophotonics, Nanyang Technological University, 50 Nanyang Avenue, Singapore 639798, Singapore and LUMINOUS! Centre of Excellence for Semiconductor Lighting \& Displays, Nanyang Technological University, 50 Nanyang Avenue, Singapore 639798, Singapore}

\date{\today}

\begin{abstract}
Inverted type-I heterostructure core/crown quantum rings (QRs) are quantum-efficient luminophores, whose spectral characteristics are highly tunable. Here, we study the optoelectronic properties of type-I core/crown CdS/CdSe QRs in the zincblende phase -- over contrasting lateral size and crown width. For this we inspect their strain profiles, transition energies, transition matrix elements, spatial charge densities, electronic bandstructure, band-mixing probabilities, optical gain spectra, maximum optical gains and differential optical gains. Our framework uses an effective-mass envelope function theory based on the 8-band \textbf{\textit{k$\cdot$p}} method employing the valence force field model for calculating the atomic strain distributions. The gain calculations are based on the density-matrix equation and take into consideration the excitonic effects with intraband scattering. Variations in the QR lateral size and relative widths of core and crown (ergo the composition) affect their energy levels, band-mixing probabilities, optical transition matrix elements, emission wavelengths/intensity, etc. The optical gain of QRs is also strongly dimension and composition dependent with further dependency on the injection carrier density causing band-filling effect. They also affect the maximum and differential gain at varying dimensions and compositions.
\end{abstract}

\maketitle

\section{\label{sec:intro}Introduction}

Heterostructure quantum rings (QRs) are a recently developed class of nanocrystals in which a traditionally used active material of lower bandgap is laterally grown around a traditionally used barrier material of higher bandgap. A typical example is the inverted Type-I CdS/CdSe core/crown colloidal QR.\cite{Lamarre15,Bose17QRing} This is a topological inversion of the popularly studied CdSe/CdS nanoplatelets (NPLs).\cite{Tessier14} Core/crown NPLs exhibit enhanced quantum confinement and rapid transfer of excitons from CdS crown to CdSe core, but their emission is mostly governed by the core dimensions. However, in QRs the emission window is extended, which originates from the excitonic recombinations in the CdSe crown, as expected from the band alignment.\cite{Lamarre15} In a QR, the laterally grown crown may be rectangular\cite{Lamarre15,Bose17QRing} or toroid-like,\cite{Fedin16} depending on the synthesis techniques. Recently, Fedin \textit{et al.} have demonstrated the colloidal synthesis of luminescent CdSe QRs via controlled etching of NPLs.\cite{Fedin16} There are other synthesis methods, such as the one-pot synthesis technique,\cite{Lamarre15} and the CdS seed technique.\cite{Delikanli2015QR} Owing to the non-trivial topology of semiconductor QRs, they exhibit interesting optoelectronic properties. They have high optical tunability and their emission characteristics can be spectrally tuned by varying the extent of lateral confinement. They exhibit high charge injection efficiency and enhanced absorption range.\cite{Delikanli2015QR} Delikanli \textit{et al.} have shown that the Type-I core/crown CdS/CdSe QRs of a particular thickness is capable of exhibiting a range of peak emission wavelengths – from that of core only CdS NPLs to that of core only CdSe NPLs having the same thickness.\cite{Delikanli2015QR} These CdS/CdSe QRs have already found application as phosphors for color conversion in white light LEDs.\cite{Delikanli2015QR} They are also potential candidates for magneto-optical device applications (using Aharonov-Bohm effect).\cite{Fedin16} Colloidal QRs could also be used as charge separator for solar cells and for light harvesting, light-emitting diode, active medium for laser and luminescent probe for biomedical imaging.\cite{Lamarre15,guzelturk14} QRs are also intriguing from a fundamental theoretical physics point-of-view and for light-matter interactions. To understand the physics of QRs for such applications, it is important to study how-and-why its optoelectronic properties can be tailored by varying the QR core and crown dimensions. In this work, we will study the electronic structure along with strain profile and charge density, plus the optical gain characteristics for QRs of varying core and crown widths.

	\begin{figure*}[t]
		\centering
		\subfloat[A 3D atom-by-atom view of a typical QR (\textit{example}: 3 ML CdS/CdSe QR -- of CdS core and CdSe crown). Red spheres are Cd atoms, yellow spheres are Se atoms and dark blue spheres are S atoms.]{\includegraphics[trim={12cm 6.8cm 11cm 8cm},width=0.52\textwidth]{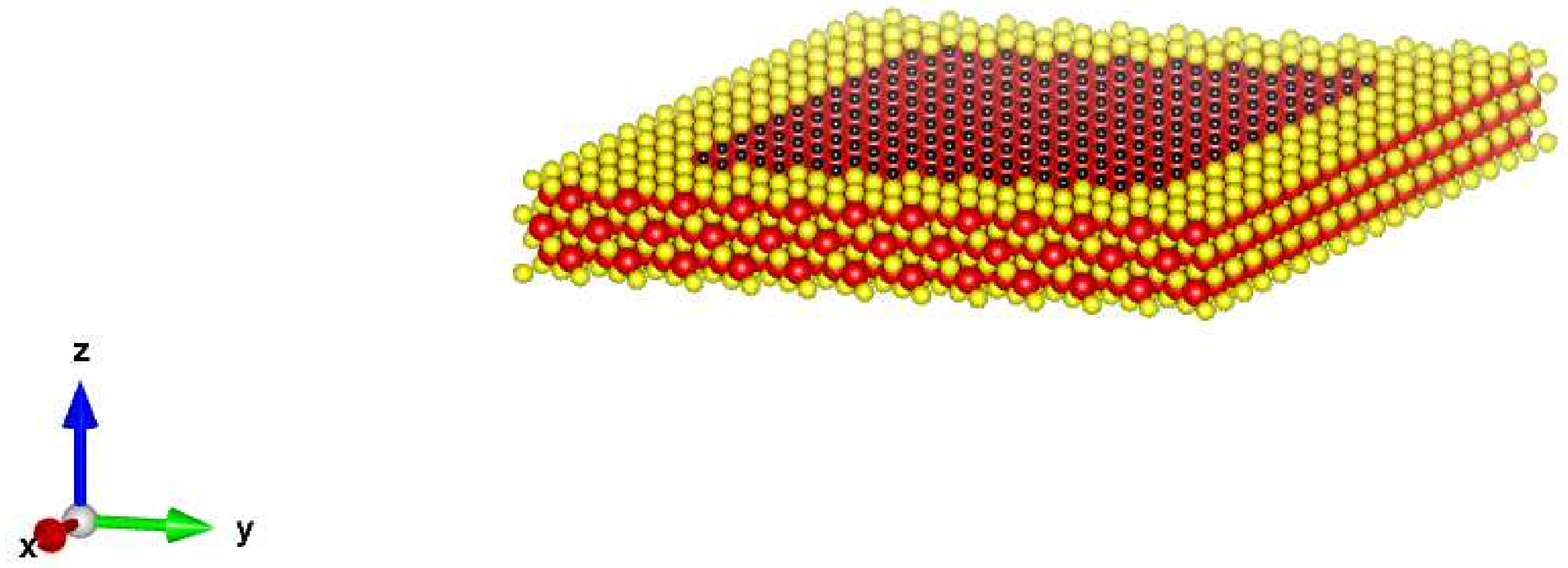}\label{subfig:qr-3d}}
		\subfloat[Schematic of a typical inverted type-I core/crown CdS/CdSe quantum ring of dimensions \textit{a}$\times$\textit{b}$\times$\textit{t} and crown width \textit{d}.]{\includegraphics[width=0.42\textwidth]{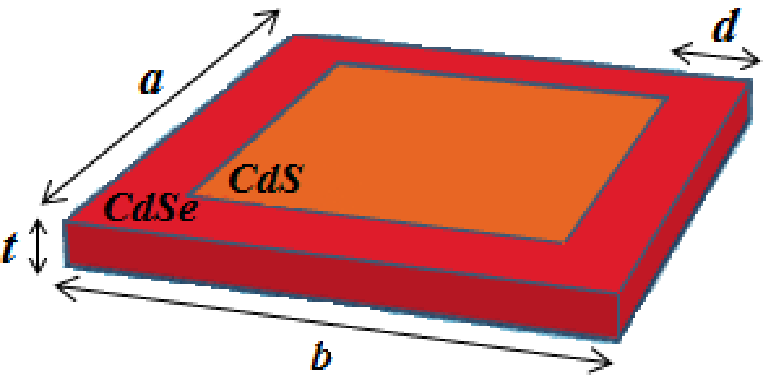}\label{subfig:qr-schematic}}
		\caption{(a: \textit{left}) 3D atom-by-atom view, and (b: \textit{right}) schematic representation of a typical QR.}
		\label{fig:qr-3d-schematic}
	\end{figure*}

\section{\label{sec:theory}Theoretical Framework}

We have studied inverted type-I core/crown CdS/CdSe QRs of 5 monolayer (ML) thickness in the zincblende (ZB) phase for varying core and crown dimensions, assuming that they are colloidally synthesized.\cite{Lamarre15,Fedin16,Delikanli2015QR} The dielectric ligand environment in the colloidal solution passivates the surface electronic states, where the QRs exist in a free-standing form. Fig.\ \ref{fig:qr-3d-schematic} (a) demonstrates a 3D atom-by-atom view of a typical QR -- showing an example of 3 ML CdS core and CdSe crown forming a rectangular type-I core/crown CdS/CdSe QR. Fig.\ \ref{fig:qr-3d-schematic} (b) shows the schematic of a typical QR of dimensions $a\times b\times t$. The width of the CdSe crown (ring) is $d$.

\subsection{\label{subsec:ebs-og}Electronic Bandstructure Calculation}

The electronic bandstructure of materials can be extrapolated from the knowledge of a set of material parameters using the \textbf{\textit{k$\cdot$p}} method, which is usually evaluated at a single point of the reciprocal space.\cite{Voon,Marconcini11} The \textbf{\textit{k$\cdot$p}} method can be used to obtain the bandstructure of a range of bulk or quantum confined semiconductor types including diamond, zincblende (ZB) and wurtzite.\cite{Voon} Moreover, it can also be extended to study graphene, carbon nanotubes and graphene nanoribbons.\cite{Marconcini11}

In our work, for every inverted type-I core/crown CdS/CdSe QR sample, we have used an effective mass envelope function theory approach based on the 8-band \textbf{\textit{k$\cdot$p}} method accounting for the nonparabolicity of the coupled conduction band (CB) and valence band (VB) simultaneously, including the orbit-splitting bands to obtain its electronic structure by solving for its eigenenergy in the $\Gamma$-point vicinity of the Brillouin zone.\cite{chen09} The 8-band Hamiltonian, $\mathscr{H}_{8\times8}$ as given by Eq.\ \eqref{eq:H-main} is represented in the Bloch function basis of $\vert S\rangle\uparrow$, $\vert1,+1\rangle\uparrow$, $\vert1,0\rangle\uparrow$, $\vert1,-1\rangle\uparrow$ and 
$\vert S\rangle\downarrow$, $\vert1,+1\rangle\downarrow$, $\vert1,0\rangle\downarrow$, $\vert1,-1\rangle\downarrow$. Here $\vert1,\pm1\rangle\uparrow\downarrow=\left(\vert X\rangle\pm i\vert Y\rangle\right)/\sqrt{2}\uparrow\downarrow$ and $\vert1,0\rangle\uparrow\downarrow=\vert Z\rangle\uparrow\downarrow$.

\begin{widetext}
	\begin{equation}\label{eq:H-main}
	\mathscr{H}_{8\times8}=\left[ \begin{array}{cccccccc}
	h_{11} & h_{12} & h_{13} & h_{14} & 0 & 0 & 0 & 0 \\
	& h_{22} & h_{23} & h_{24} & 0 & 0 & 0 & 0 \\
	&   & h_{33} & h_{34} & 0 & 0 & 0 & 0 \\
	&   &   & h_{44} & 0 & 0 & 0 & 0 \\
	&   &   &   & h_{11} & h_{12} & h_{13} & h_{14} \\
	&   &   &   &   & h_{22} & h_{23} & h_{24} \\
	&   &   &   &   &   & h_{33} & h_{34} \\
	&   &   &   &   &   &   & h_{44} \\
	\end{array} \right]
	+\dfrac{\Delta_{so}}{3}
	\left[ \begin{array}{cccccccc}
	0 & 0 & 0 & 0 & 0 & 0 & 0 & 0 \\
	0 & 0 & 0 & 0 & 0 & 0 & 0 & 0 \\
	0 & 0 & -1 & 0 & 0 & -\sqrt{2} & 0 & 0 \\
	0 & 0 & 0 & -2 & 0 & 0 & \sqrt{2} & 0 \\
	0 & 0 & 0 & 0 & 0 & 0 & 0 & 0 \\
	0 & 0 & -\sqrt{2} & 0 & 0 & -2 & 0 & 0 \\
	0 & 0 & 0 & \sqrt{2} & 0 & 0 & -1 & 0 \\
	0 & 0 & 0 & 0 & 0 & 0 & 0 & 0 \\
	\end{array} \right]
	+\Phi_{\text{QR}}
	\end{equation}
\end{widetext}

All the terms independent or linearly/quadratically dependent on the wavevector are contained in the first Hamiltonian matrix, with $h_{ij}=h_{ji}^\ast$. The valence band spin-orbit coupling\cite{chen09} is accounted by the second Hamiltonian matrix; and the QR confining potential from the dielectric ligand environment is $\Phi_{\text{QR}}$. Expressions of the Hamiltonian elements ($h_{ij}$) are given below.

\begin{subequations}
	\label{eqn:hamil-elements}
	
	\begin{equation}\label{eq:h11}
	h_{11}=E_g-\dfrac{\hbar^2}{2m_0}\cdot\gamma_c\left(k_x^2+k_y^2+k_z^2\right)+a_c\left[tr\left(\varepsilon\right)\right]
	\end{equation}
	
	\begin{equation}
	h_{12}=\frac{i\hbar\sqrt{E_p}\left(k_x^\prime+ik_y^\prime\right)}{2\sqrt{m_0}}
	\end{equation}
	
	\begin{equation}
	h_{13}=i\hbar\sqrt{\frac{E_p}{2m_0}}k_z^\prime
	\end{equation}
	
	\begin{equation}
	h_{14}=\frac{i\hbar\sqrt{E_p}\left(k_x^\prime-ik_y^\prime\right)}{2\sqrt{m_0}}
	\end{equation}
	
	\begin{eqnarray}
	h_{22}=h_{44}=-\dfrac{\hbar^2}{2m_0}\left[\dfrac{L^\prime+M^\prime}{2}\left(k_x^2+k_y^2\right)+M^\prime k_z^2\right] \nonumber\\+ a_v\left[tr\left(\varepsilon\right)\right] + \frac{b}{2}\left[tr\left(\varepsilon\right)-2\varepsilon_{zz}\right]\text{\hspace{10mm}}
	\end{eqnarray}
	
	\begin{eqnarray}
	h_{23}=h_{34}=-\dfrac{\hbar^2}{2m_0}\left[\dfrac{N^\prime\left(k_x-ik_y\right)k_z}{\sqrt{2}}\right]\nonumber\\+\sqrt{6}d\left(\varepsilon_{xz}-i\varepsilon_{yz}\right)
	\end{eqnarray}		
	
	\begin{eqnarray}
	h_{24}=-\dfrac{\hbar^2}{2m_0}\left[\dfrac{L^\prime-M^\prime}{2}\left(k_x^2-k_y^2\right)-iN^\prime k_xk_y\right] \nonumber\\+ 
	\frac{3b}{2}\left(\varepsilon_{xx}-\varepsilon_{yy}\right)-i\sqrt{12}d\varepsilon_{xy}\text{\hspace{17mm}}
	\end{eqnarray}
	
	\begin{eqnarray}\label{eq:h33}
	h_{33}=-\dfrac{\hbar^2}{2m_0}\left[M^\prime\left(k_x^2+k_y^2\right)+L^\prime k_z^2\right] \nonumber\\+ a_v\left[tr\left(\varepsilon\right)\right] + b\left[3\varepsilon_{zz}-tr\left(\varepsilon\right)\right]\text{\hspace{0.5mm}}
	\end{eqnarray}

    \begin{equation}
	\left(
	\begin{array}{c}
	k_x^\prime \\ k_y^\prime \\ k_z^\prime
	\end{array}
	\right) =\left(
    \begin{array}{ccc}
    1-\varepsilon_{xx} & -\varepsilon_{xy} & -\varepsilon_{xz}\\
    -\varepsilon_{yx} & 1-\varepsilon_{yy} & -\varepsilon_{yz}\\
    -\varepsilon_{zx} & -\varepsilon_{zy} & 1-\varepsilon_{zz}\\
    \end{array}\right)
	\left(
	\begin{array}{c}
	k_x \\ k_y \\ k_z
	\end{array}
	\right) 
\end{equation}
	
\end{subequations}

Here $k_x^\prime$, $k_y^\prime$, $k_z^\prime$ are the modified wavevectors and $\varepsilon$ is the strain tensor matrix. $L^\prime$, $M^\prime$ and $N^\prime$ are the modified Luttinger parameters, derived from the Luttinger-Kohn effective mass parameters $\gamma_1$, $\gamma_2$, $\gamma_3$.\cite{bose16,Bose17NS} The expansion of the eight-dimensional hole and electron envelope wavefunctions using plane waves is done using the method described in our previous work.\cite{bose16,Bose17NS}

The strain tensor terms in the Hamiltonian influence the wavefunctions and electronic bandstructures. The following two methods are commonly used for strain calculations: While the (\textit{i}) atomistic valence force field (VFF) model is computationally more demanding, it provides details on inter-atomic scales accounting for the full crystal lattice structure. On the other hand, the (\textit{ii}) continuum models (CM) provides computationally fast and accurate results for nanostructures with dimensions much larger than lattice constant,\cite{song17_top_ins,rev_Stier99} but fails to produce the polarization dependent anisotropy in electron-hole transition matrix elements.\cite{rev_Sengupta14} Lazarenkova \textit{et al.}\cite{rev_Lazarenkova04} have observed that the CM is inadequate in describing the strain profile in heterostructures with large lattice mismatch (7\% in ref.\ \citen{rev_Lazarenkova04}). While not universally true, this particularly holds when the size of the nanostructure has comparable dimensions with the reticular constant. In our case, the CdS-CdSe heterostructures lattice mismatch is comparable at $\sim$4\%, where the anharmonicity of the interatomic potential becomes significant. In the VFF model, the anharmonicity is added to the inter-atomic potential by utilizing distance-dependent (bond-stretching) and angle-dependent (bond-bending) force constants ($\alpha$ and $\beta$).\cite{rev_Lazarenkova04} The anharmonicity is a non-linear dependence, and non-linearity is inherent in the atomistic description, which is not captured well by CM as explained in ref. \citen{rev_Sengupta14}. There are several comprehensive works by Renka,\cite{renka88} Gullett \textit{et al.}\cite{gullett08} and Barettin \textit{et al.}\cite{barettin12} describing the computational methods for electromechanical fields in self-assembled nanostructures and methods to calculate the deformation gradient tensor and strain tensors for atomistic simulations. The approach of Renka relies on the quadratic Shepard method for bivariate interpolation, while that of Gullett \textit{et al.} is a kinematical algorithm for the construction of strain tensors from atomistic simulation data; both giving similar results. However, in the work of Barettin \textit{et al.} they discuss both CM and VFF methods. In this work, we utilize the VFF model to calculate the strain tensor used in the \textbf{\textit{k$\cdot$p}} Hamiltonian given by Eq.\ \eqref{eq:H-main}.\footnote{Usually the Pseudopotential Method and Empirical Tight Binding atomistic bandstructure calculations uses VFF strain model. But in the \textbf{\textit{k$\cdot$p}} method both VFF and CM strain models can be used, and the choice of the most appropriate strain model is application specific and depends on the nanostructure in question.\cite{rev_Sengupta14}} The total VFF strain energy is given by\cite{bose14b,Bose17NS}

\begin{eqnarray} 
	E_{\text{VFF}}=\sum\limits_{i(j)}^{ }\dfrac{3\alpha_{ij}}{16d_{0,ij}^2}\bigg(\vert\textbf{r}_i-\textbf{r}_j\vert^2-d_{0,ij}^2\bigg)^2+\sum\limits_{i(j,k)}^{ }\dfrac{3\beta_{jik}}{8d_{0,ij}d_{0,ik}} \nonumber\\
	\bigg(\vert\textbf{r}_i-\textbf{r}_j\vert\vert\textbf{r}_i-\textbf{r}_k\vert-\cos\hat{\theta}_{jik}\cdot d_{0,ij}d_{0,ik}\bigg)^2 \text{\hspace{6mm}}
	\label{eqn:vff}
\end{eqnarray}

\noindent where the atoms in the crystal lattice are identified by the indices \textit{i}, \textit{j} and \textit{k}. $d_{0,ij}$ is the ideal atomic bond-length between the $i^{th}$ and $j^{th}$ atoms, with $4 d_{0,ij}=\sqrt{3} a_{0,ij}$ relating it to $a_{0,ij}$, the lattice constant.\cite{bose14a} External factors cause deviations in the bond-length from the ideal $d_{0,ij}$, and the bond-distance between the $i^{th}$ and $j^{th}$ atoms in the system studied is given by $\vert\textbf{r}_i-\textbf{r}_j\vert$, where $\textbf{r}_i$ is the position vector of the $i^{th}$ atom. The bond-stretching force constant of the $i-j$ bond is given by $\alpha_{ij}$. The ideal bond-angle of the bond among the $j^{th}$, $i^{th}$ and $k^{th}$ atoms (vertexed at $i^{th}$ atom) is given by $\hat{\theta}_{jik}$ ($=\frac{1}{3}$ for ZB structures). The bond-bending force constant of the $j-i-k$ bond-angle is given by $\beta_{jik}$. The material parameters used in this work are from ref. \citen{bose16}.

\subsection{\label{subsec:opt-prop-calc}Optical Properties Calculation}

On the basis of the density-matrix theory equation, we calculate the optical gain spectra of the QR considering excitonic effects as a sum of the contributions from excitonic bound states, $\mathscr{G}_{sp}^{ex,b}$ and band-to-band continuum-states, $\mathscr{G}_{sp}^c$. The excitonic bound state contributions is given by\cite{micallef93,herbert92} 

\begin{subequations}
	\begin{eqnarray}\label{Gspb}
	\mathscr{G}_{sp}^{ex,b}\left(\hbar\omega\right)=\dfrac{C_0}{t}\sum\limits_{c,v}\vert\Psi_{1s}^{cv}\left(0\right)\vert^2\vert\mathscr{P}_{cv}\vert^2\vert {I}_{cv}\vert^2\times\nonumber\\
	\left(f_c-f_v\right)\mathscr{L}(\hbar\omega-\hbar\omega_{cv}-\hbar\omega_b)\text{\hspace{10mm}}
	\end{eqnarray}
	\begin{equation}
	\Psi_{1s}\left(x\right)=\frac{4\beta}{a_{B}\sqrt{2\pi}}e^{-2x\beta/a_B}
	\end{equation}
	\begin{equation}
	\hbar\omega_b=-4\beta^2R_y
	\end{equation}
\end{subequations}

\noindent And the band-to-band continuum state contributions is given by\cite{herbert92,Chuang} 

\begin{subequations}
	\begin{eqnarray}\label{Gspc}
	\mathscr{G}_{sp}^c\left(\hbar\omega\right)=\dfrac{C_0}{V}\sum\limits_{c,v}\vert\mathscr{P}_{cv}\vert^2\left(f_c-f_v\right)\times\nonumber\\
	S_{2D}\left(\hbar\omega-\hbar\omega_{cv}\right)\mathscr{L}(\hbar\omega-\hbar\omega_{cv})\text{\hspace{10mm}}
	\end{eqnarray}
	\begin{equation}
	S_{2D}\left(\hbar\omega-\hbar\omega_{cv}\right)=\frac{2}{1+\text{exp}\left(-2\pi\sqrt{R_y/(\hbar\omega-\hbar\omega_{cv})}\right)}
	\end{equation}
\end{subequations}

\noindent where $C_0=\dfrac{\pi e^2}{n_rc\varepsilon_0 m_0^2\omega}$, and symbols (\textit{e}, $n_r$, \textit{c}, $\varepsilon_0$, $m_0$) have standard physical meanings. $t$ is the QR thickness, and \textit{V} is the QR volume in real space. The overlap integral between the electron and hole wavefunctions along the transverse direction is given by ${I_{cv}}=\braket{\Psi^{E(c)}\left(z\right)|\Psi^{H(v)}\left(z\right)}$. The 1S exciton envelope function is given by $\Psi_{1s}\left(x\right)$ where \textit{x} is the relative distance between the electron and hole along the transverse direction in the NPL. $\hbar\omega_b$ is the 1S exciton binding energy and $\beta$ is a variational parameter, taken to be 1 for two-dimensional structures. $R_y=m_re^4/32\pi^2\varepsilon_0^2\varepsilon_r^2\hbar^2$ and $a_B=4\pi\varepsilon_0\varepsilon_r\hbar^2/m_re^2$ are the excitonic Rydberg energy and excitonic Bohr radius respectively, where $\varepsilon_r$ is the relative permittivity and $m_r$ is the reduced mass of the electron-hole pair: $m_r^{-1}=m_e^{-1}+m_h^{-1}$. $S_{2D}\left(\hbar\omega-\hbar\omega_{cv}\right)$ is the two-dimensional Sommerfeld enhancement factor.\cite{Chuang} $f_c$ and $f_v$ are the Fermi-Dirac distributions for the electrons and
holes in the CB and VB respectively.\cite{fan16} The gain is proportional to $(f_c-f_v)$, also known as the Fermi factor; and is maximum when $f_c=1$ and $f_v=0$.\cite{bose16} Note that the Fermi factor diminishes to 0 as $E_{cv}$ approaches towards the quasi-Fermi level separation $\Delta E_{F}$, beyond which the gain is negative (absorption occurs). The gain spectra of the QR must however consider the transition energy broadening (dephasing) which is accounted by the term $\mathscr{L}_{cv}(\hbar\omega-\lambda)$ in Eq.\ \ref{Gspb} and \ref{Gspc}, and expressed as\cite{Chuang}

\begin{equation}
\mathscr{L}_{cv}(\hbar\omega-\lambda)=\dfrac{1}{\pi}\dfrac{\hbar/\tau_{s}}{\left(\hbar\omega-\lambda\right)^2+\left(\hbar/\tau_{s}\right)^2}
\end{equation}  

\noindent where $\tau_{s}$ is the intraband relaxation time and $2\hbar/\tau_{s}$ is the FWHM. The optical transition matrix element (TME) is given by $\vert\mathscr{P}_{cv}\vert^2$, which quantitatively measures the stimulated transition strengths between hole- and electron-subband.\cite{sugawara-book99} Using the electron- and hole-wavefunctions ($\Psi_{c,\textbf{\text{k}}}$ and $\Psi_{v,\textbf{\text{k}}}$), and the momentum operator, \textbf{p} it can be calculated along \textit{x}, \textit{y} and \textit{z} as $\mathscr{P}_{cv,i}=\bra{\Psi_{c,\textbf{\text{k}}}}\textbf{\text{e}}_i\cdot\textbf{\text{p}}\ket{\Psi_{v,\textbf{\text{k}}}}$, $i=x,y,z$.\cite{fan96} The transverse electric (TE) mode optical gain, polarized in the \textit{x-y} plane comes from the average of $\mathscr{P}_{cv,x}$ and $\mathscr{P}_{cv,y}$ (TME along the \textit{x} and \textit{y} directions), while the transverse magnetic (TM) optical gain, polarized in the \textit{z} direction comes from $\mathscr{P}_{cv,z}$ (TME along \textit{z} direction). Direction specific expressions for $\mathscr{P}_{cv,i}$ can be found in our previous work.\cite{Bose17NS}

\section{\label{sec:res-disc}Results and Discussions}

    	\begin{figure*}[t]
		\centering
\includegraphics[width=0.99\textwidth]{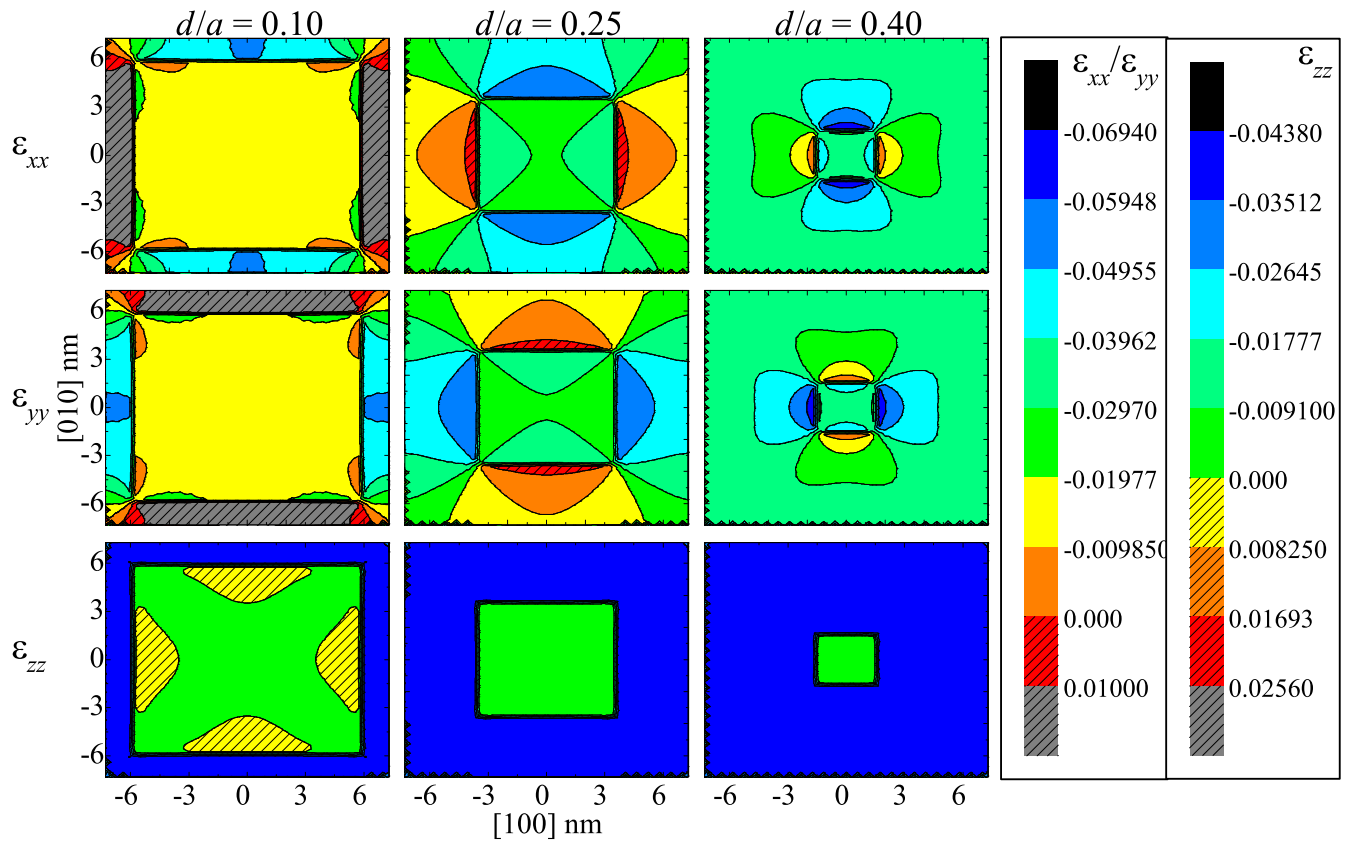}
		\caption{Strain profile distribution ($\varepsilon_{\textit{xx}}$, $\varepsilon_{\textit{yy}}$, $\varepsilon_{\textit{zz}}$) in the (001) plane of the inverted type-I CdS/CdSe QRs of lateral size 15 nm, and thickness 5 ML, with varying crown width : QR width (\textit{d}/\textit{a}) of 0.10, 0.25 and 0.40. The (001) \textit{x}-\textit{y} plane is cut horizontally along the center ($z=0$). Warmer colors (reddish) with oblique shading represent tensile strain, while cooler colors (greenish/bluish) without shading represent compressive strain. Strain component and (\textit{d}/\textit{a}) ratios are indicated. Color chart shows strain values.}
		\label{fig:STR_all}
	\end{figure*}
    
\subsection{\label{subsec:elec-Bstr}Electronic Bandstructure and Properties}

The lateral size, core width and crown width of the inverted type-I core/crown CdS/CdSe QRs are varied to investigate their influence on the strain profile, TMEs, charge densities and electronic bandstructure. Here we have considered only 5 ML thick square QRs i.e. $a=b$ for the convenience of study and understanding.

Fig.\ \ref{fig:STR_all} shows the strain profile distribution ($\varepsilon_{\textit{xx}}$, $\varepsilon_{\textit{yy}}$, $\varepsilon_{\textit{zz}}$) in the (001) plane of the CdS/CdSe QRs of lateral size 15 nm, and thickness 5 ML, with varying crown width : QR width (\textit{d}/\textit{a}) of 0.10, 0.25 and 0.40. The (001) \textit{x}-\textit{y} plane is cut horizontally along the center ($z=0$). Each row shows a particular strain component: $\varepsilon_{\textit{xx}}$ in row 1, $\varepsilon_{\textit{yy}}$ in row 2 and  $\varepsilon_{\textit{zz}}$ in row 3; while each column stands for a particular (\textit{d}/\textit{a}) ratio: 0.10, 0.25 and 0.40. The strain components and the (\textit{d}/\textit{a}) ratios are indicated. Color chart shows strain values.

Warmer colors (reddish) with oblique shading represent tensile strain, while cooler colors (greenish/bluish) without shading represent compressive strain. The $\varepsilon_{\textit{xx}}$ strain profile is anisotropic in the (001) plane, and so is $\varepsilon_{\textit{yy}}$. For both $\varepsilon_{\textit{xx}}$ and $\varepsilon_{\textit{yy}}$ there is compressive strain within the CdS core which increases for thicker crowns (rings) as the (\textit{d}/\textit{a}) ratio rises. At the core/crown boundary there is an abrupt change in the strain profile. For $\varepsilon_{\textit{xx}}$, as we move from the core to the crown, the strain type changes to tensile along the [100] direction, while along the [010] direction the extent of compressive strain increases. The magnitude of  tensile (compressive) strain along the [100] ([010]) direction at the core/crown boundary also depends on the (\textit{d}/\textit{a}) ratio. It is higher for thinner crowns and falls as the crown width increases.

Here $\varepsilon_{\textit{xx}}$ and $\varepsilon_{\textit{yy}}$ can be correlated, as $\varepsilon_{\textit{xx}}$ along [010] is identical to $\varepsilon_{\textit{yy}}$ along [100]; and $\varepsilon_{\textit{yy}}$ in the [010] direction is identical to $\varepsilon_{\textit{xx}}$ in the [100] direction. Therefore, for $\varepsilon_{\textit{yy}}$, as we move from the core to the crown, the strain type changes to tensile along the [010] direction, while along the [100] direction the extent of compressive strain increases. The dependency of the strain magnitude on the (\textit{d}/\textit{a}) ratio is similar to that of $\varepsilon_{\textit{xx}}$.

For $\varepsilon_{\textit{zz}}$, the strain pattern is predominantly compressive, and comparatively higher in the crown compared to the core. However for QRs with thinner crown, there is marginal tensile strain within the core. The calculation of strain profile is important as it affects the Hamiltonian matrix [Eq.\ \eqref{eq:H-main}] and thus the energy dispersion characteristics and electronic bandstructure.

    	\begin{figure*}[t]
		\centering
\includegraphics[width=\textwidth]{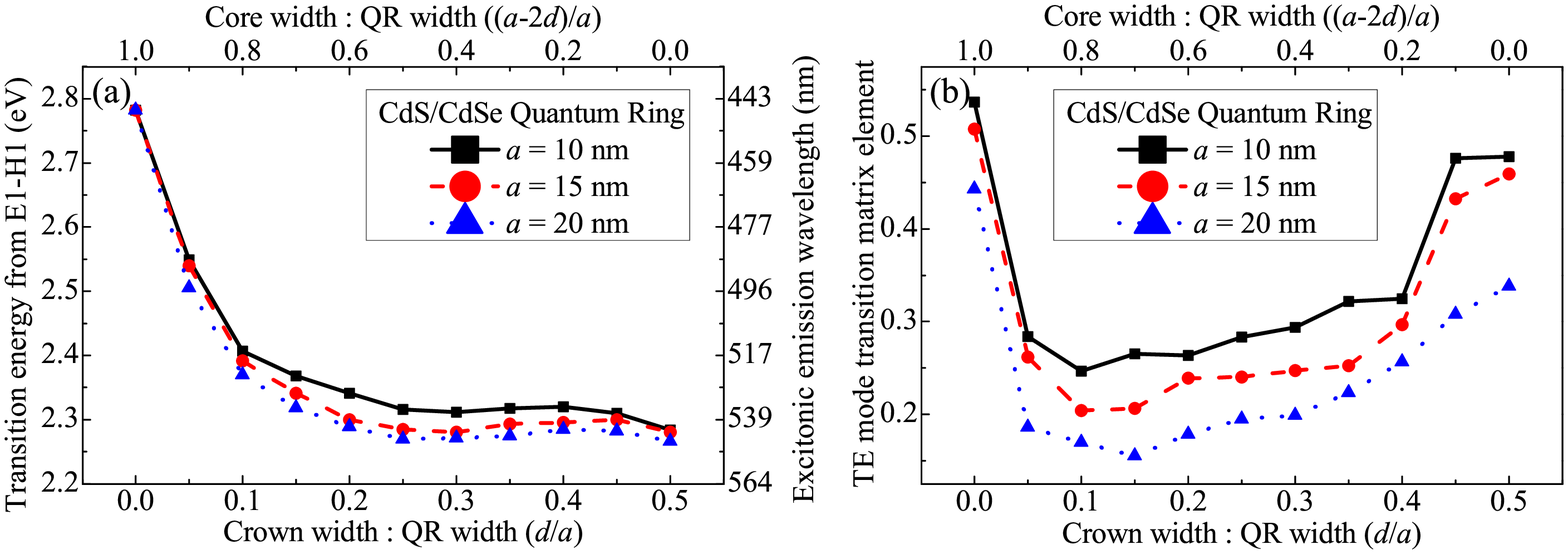}
		\caption{(a: \textit{left}) E1--H1 excitonic transition energy (photon emission wavelength), and (b: \textit{right}) Transverse electric (TE) mode transition matrix element (TME) vs. crown width : QR width (\textit{d}/\textit{a}), with fixed \textit{a} = 10, 15, 20 nm.}
        \label{fig:EN_TME_plot2}
	\end{figure*}

Now, in Fig.\ \ref{fig:EN_TME_plot2} (a) we show the excitonic transition energy (and emission wavelength) between the bottom of the conduction band (E1) to the top of the valence band (H1) i.e. E1--H1 for varying QR lateral size (\textit{a}) and crown width : QR width (\textit{d}/\textit{a}), for 5 ML thick QRs. For any QR having a given lateral size, as the (\textit{d}/\textit{a}) ratio rises from 0 to 0.5, the emission energy falls from 2.78 eV to 2.28 eV, since the QR transforms from pure CdS NPL (\textit{d}/\textit{a} $=0$) to pure CdSe NPL (\textit{d}/\textit{a} $=0.5$). A higher (\textit{d}/\textit{a}) ratio means that the QR has a higher fraction of CdSe over CdS -- the former having a lower bandgap compared to the latter, and hence the pattern. However, the transition energy pattern is not truly monotonic, but has a concavity due to the influence of optical bowing coefficient of CdS$_x$Se$_{1-x}$ = 0.28 eV,\cite{Wei00bowing} which induces the minima to occur for an intermediate composition, ergo an intermediate (\textit{d}/\textit{a}) ratio.

With an increase in the QR lateral size, there is a red-shift in the transition energy. However, when the QR transforms into pure CdS (\textit{d}/\textit{a} $=0$) or pure CdSe (\textit{d}/\textit{a} $=0.5$) NPL, the significance of lateral size (\textit{a}) diminishes and the emission energy is almost identical for 10, 15 and 20 nm as expected, since the thickness for these three cases are the same, which is the primary determinant of the NPL emission energy. But for intermediate (\textit{d}/\textit{a}) ratios, as the QR crown grows, its influence comes into picture and the $\Delta$(E1--H1) between $a=10, 15, 20$ nm increases up to a point when the core is about half the QR width and decreases thereafter.

\begin{figure*}[t]
		\centering
        
		\subfloat[For \textit{d}/\textit{a} = 0.10, charge density of CdS/CdS QR of lateral size 15 nm and thickness 5 ML.]{
			\label{subfig:WFS3D_15_80_10}
			\includegraphics[trim={4.4cm 0.4cm 3.9cm 0.3cm},width=6.2in]{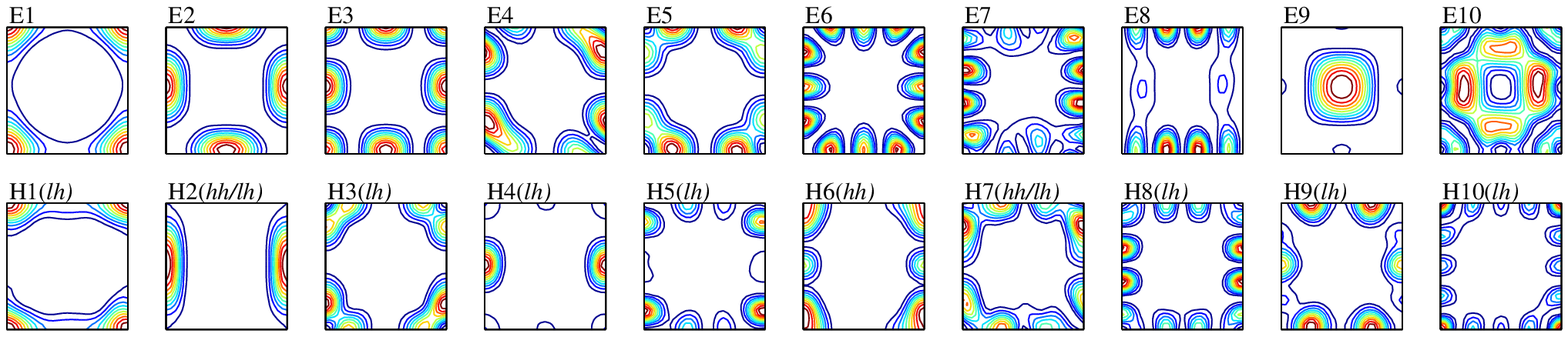} }
            
		\subfloat[For \textit{d}/\textit{a} = 0.25, charge density of CdS/CdS QR of lateral size 15 nm and thickness 5 ML.]{
			\label{subfig:WFS3D_15_50_25}
			\includegraphics[trim={4.4cm 0.4cm 3.9cm 0.3cm},width=6.2in]{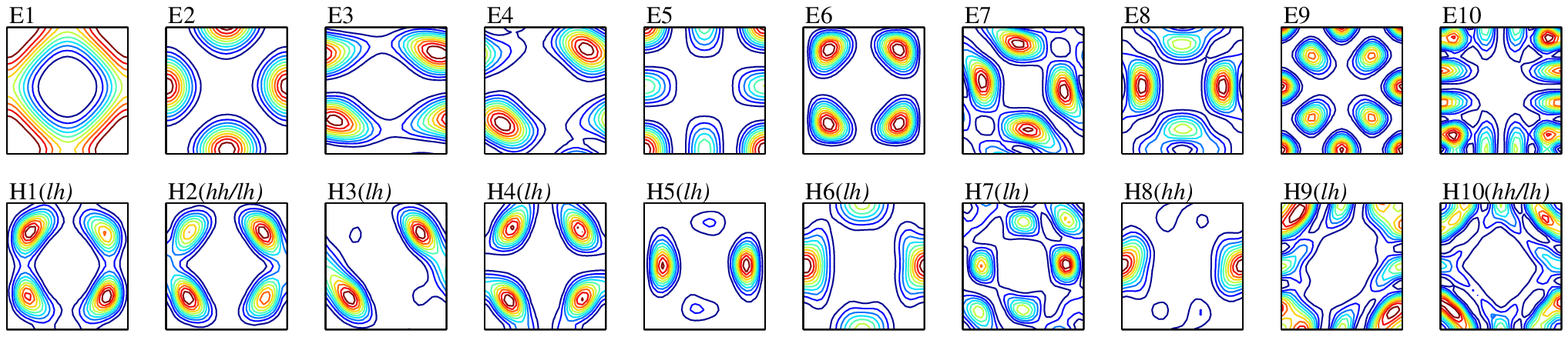} }
            
		\subfloat[For \textit{d}/\textit{a} = 0.40, charge density of CdS/CdS QR of lateral size 15 nm and thickness 5 ML.]{
			\label{subfig:WFS3D_15_20_40}
			\includegraphics[trim={4.4cm 0.4cm 3.9cm 0.3cm},width=6.2in]{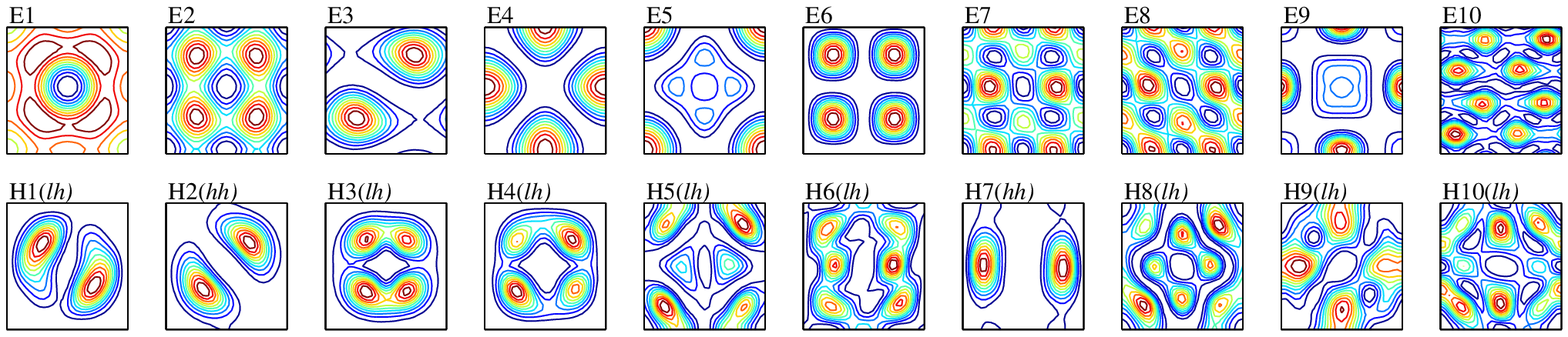} }
            
		\caption{Spatial charge density distributions of the first ten conduction (E) and valence (H) states of inverted type-I CdS/CdSe QR of lateral size 15 nm and thickness 5 ML, with varying crown width : QR width of (a: \textit{top}) \textit{d}/\textit{a} = 0.10, (b: \textit{middle}) \textit{d}/\textit{a} = 0.25 and (c: \textit{bottom}) \textit{d}/\textit{a} = 0.40. Charge densities are shown in the (001) \textit{x}-\textit{y} plane cut horizontally along the center ($z=0$). Most prominent hole-types (\textit{hh} for heavy hole and \textit{lh} for light hole) for every H state are mentioned from the band-mixing probabilities shown in Fig.\ \ref{fig:EN_Prob_all}.}
		\label{fig:WFS3D-all}
	\end{figure*}

Fig.\ \ref{fig:EN_TME_plot2} (b) shows the TE mode transition matrix element (TME) of the E1--H1 transitions, for the QR cases studied in Fig.\ \ref{fig:EN_TME_plot2} (a). For QRs that have transformed into pure CdS NPL (\textit{d}/\textit{a} $=0$) or pure CdSe NPL (\textit{d}/\textit{a} $=0.5$), the TME values are higher as both the electron and hole wavefunction are localized within the single material without any leakage. However for CdS/CdSe QR heterostructures, the TMEs diminish and is dependent on the extent of overlap between electron- and hole-wavefunctions, that are mainly localized in the CdSe crown but also leaks into the CdS core. This can be seen from Fig.\ \ref{fig:WFS3D-all} which shows the spatial charge densities of the first ten electron and hole states for CdS/CdSe QRs of lateral size 15 nm, and thickness 5 ML, with varying crown width : QR width (\textit{d}/\textit{a}) of 0.10, 0.25 and 0.40. These are the three cases that were studied in Fig.\ \ref{fig:STR_all} also. For a sufficiently low (\textit{d}/\textit{a}) ratio, such as 0.10 the extent of electron- and hole-wavefunctions overlap is limited, as can be seen from the E1 (1$^\text{st}$ conduction electron level) and H1 (1$^\text{st}$ valence hole level) charge density distribution in Fig.\ \ref{fig:WFS3D-all} (a) -- this limits the TME. As the (\textit{d}/\textit{a}) ratio increases, the CdSe crown becomes larger and the spatial electron-hole wavefunctions overlap also increases [compare E1 and H1 of frame (a) with frame (b) and (c) in Fig.\ \ref{fig:WFS3D-all}]. This results in the gradual increase in the TME with increasing (\textit{d}/\textit{a}). For a given (\textit{d}/\textit{a}), however, the TME is higher for QRs with lower lateral size as the electron- and hole-wavefunctions are more compactly packed enabling a greater overlap.

\begin{figure*}
			\centering
\includegraphics[scale=0.257, angle=270 ]{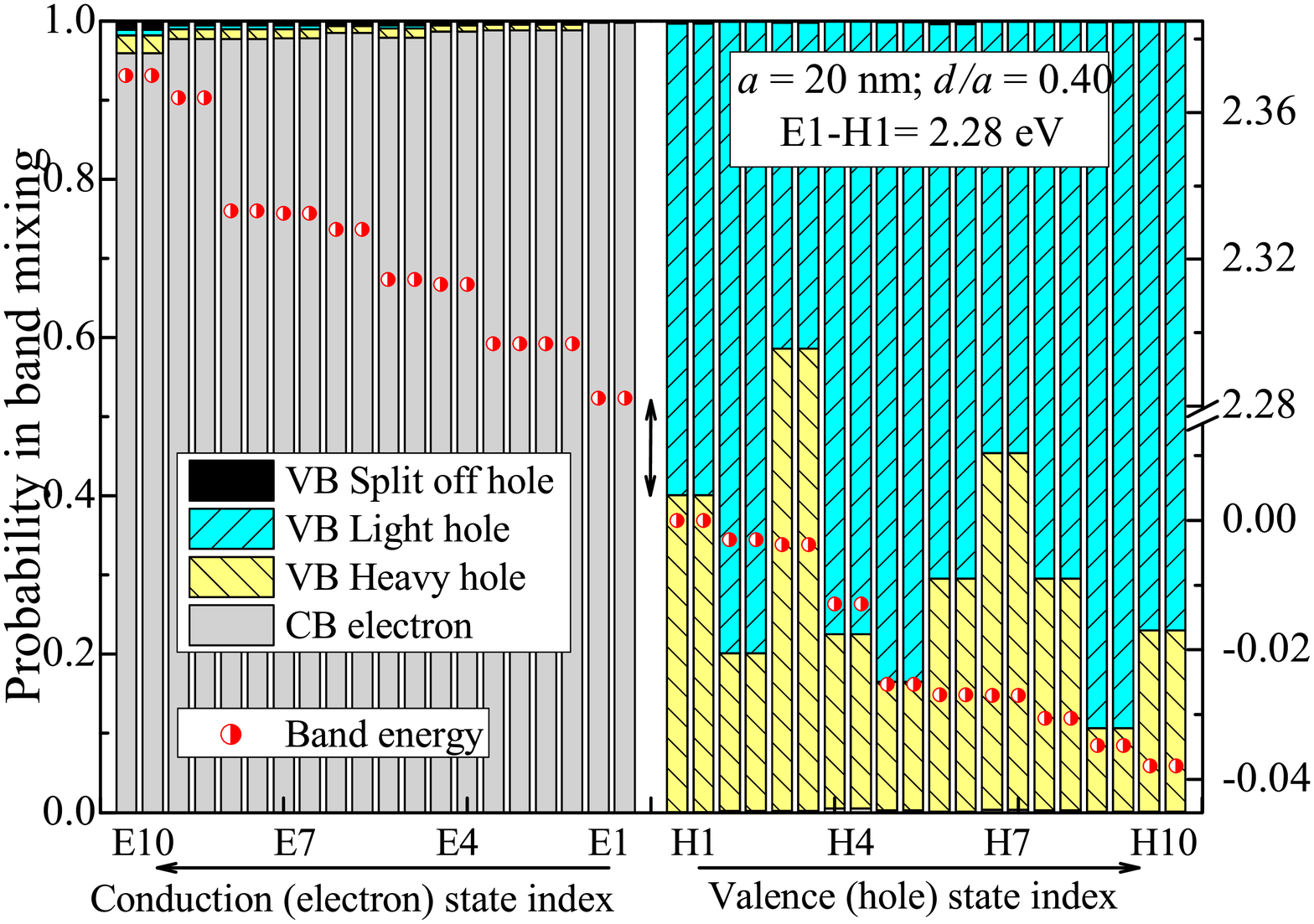}
\includegraphics[scale=0.257, angle=270 ]{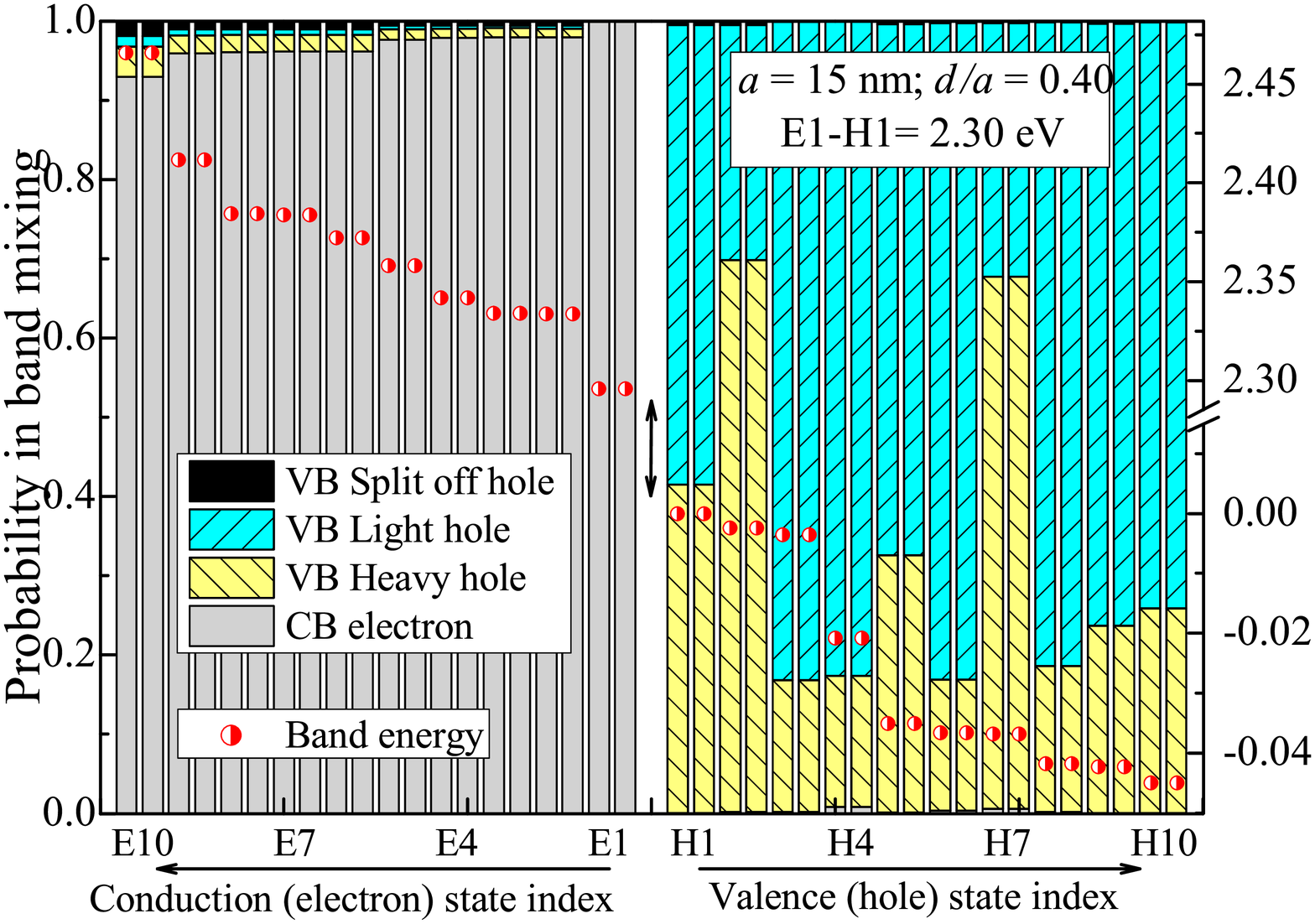}
\includegraphics[scale=0.257, angle=270 ]{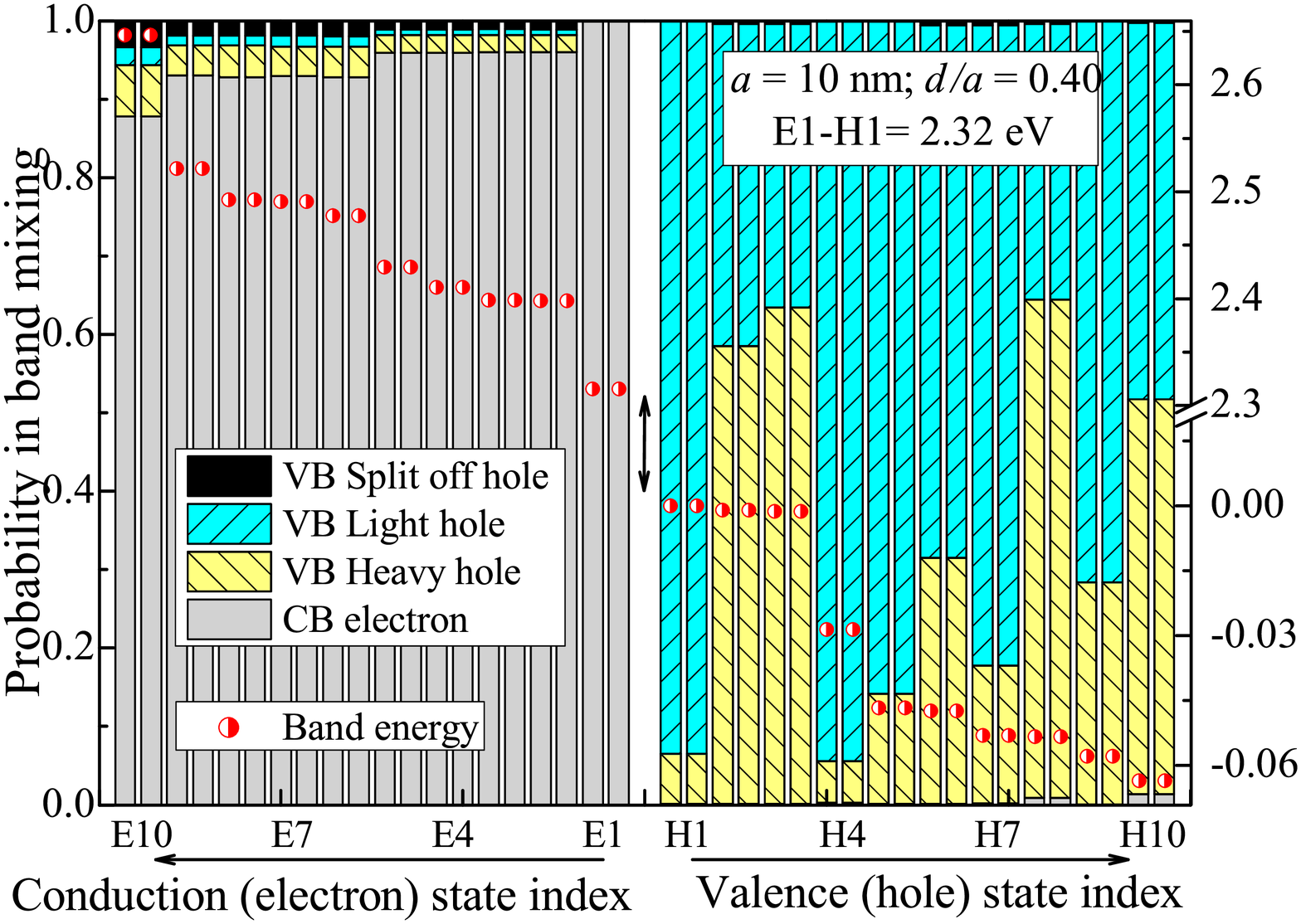}\\
\includegraphics[scale=0.252, angle=270 ]{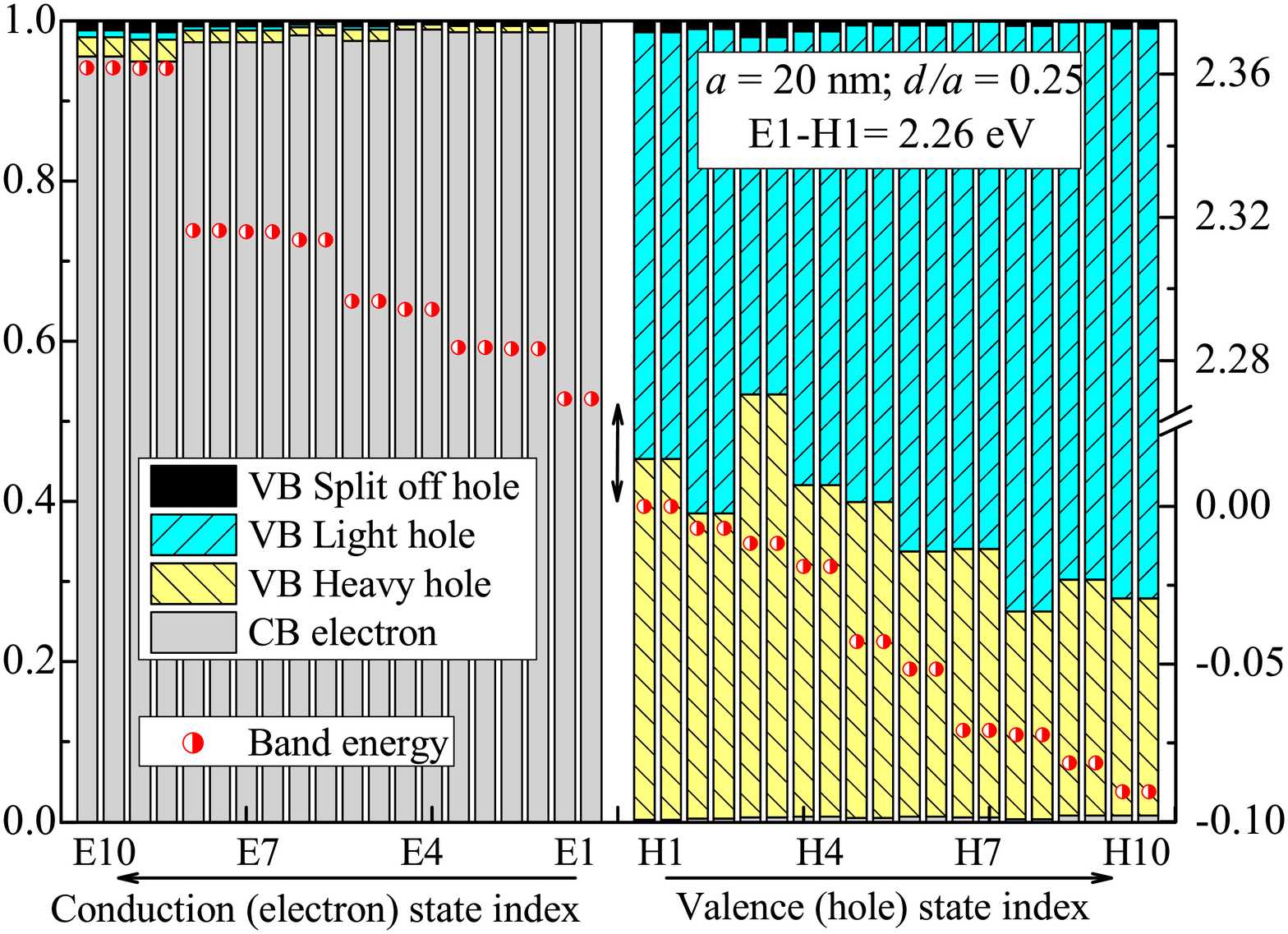}
\includegraphics[scale=0.252, angle=270 ]{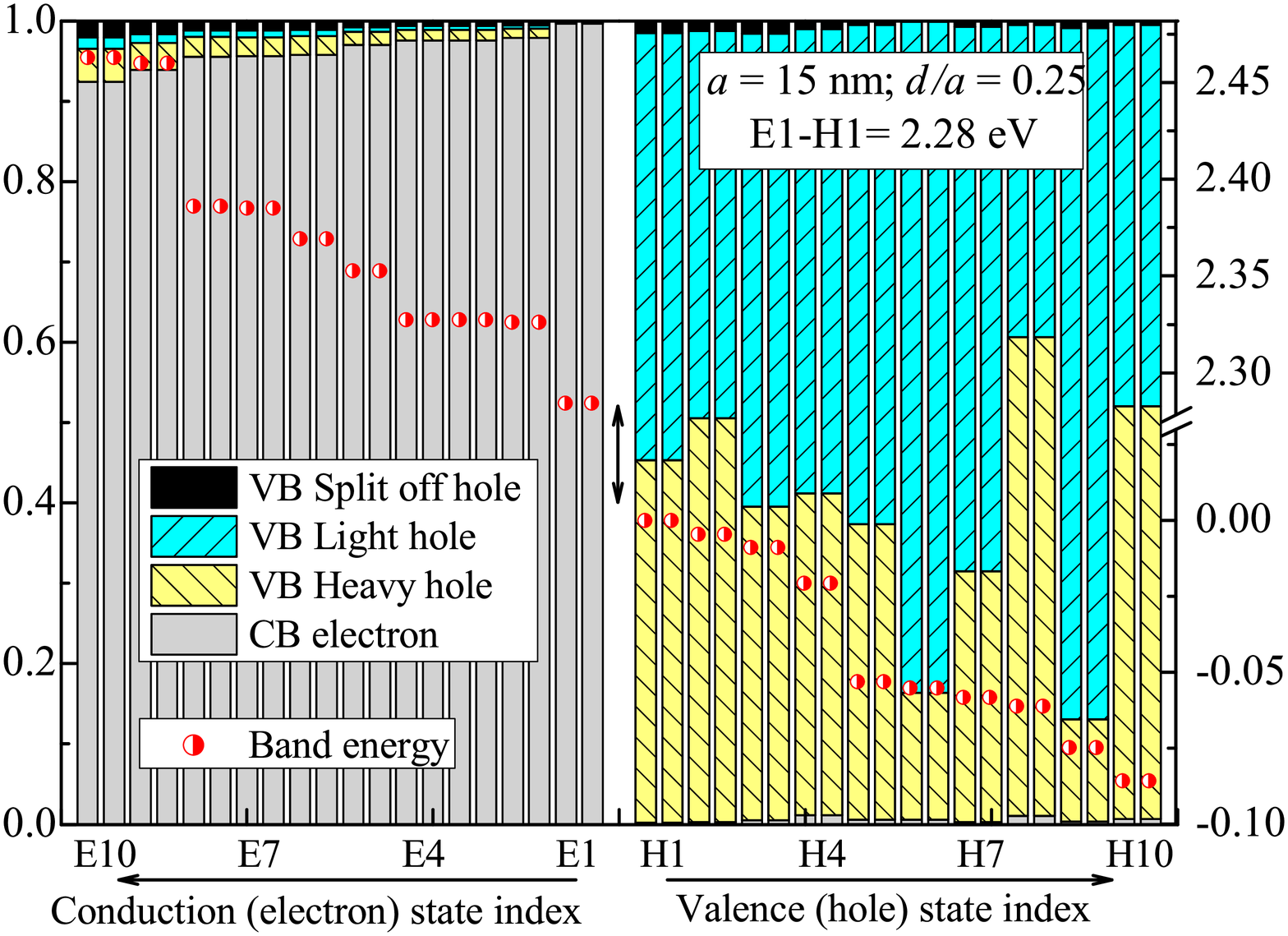}
\includegraphics[scale=0.252, angle=270 ]{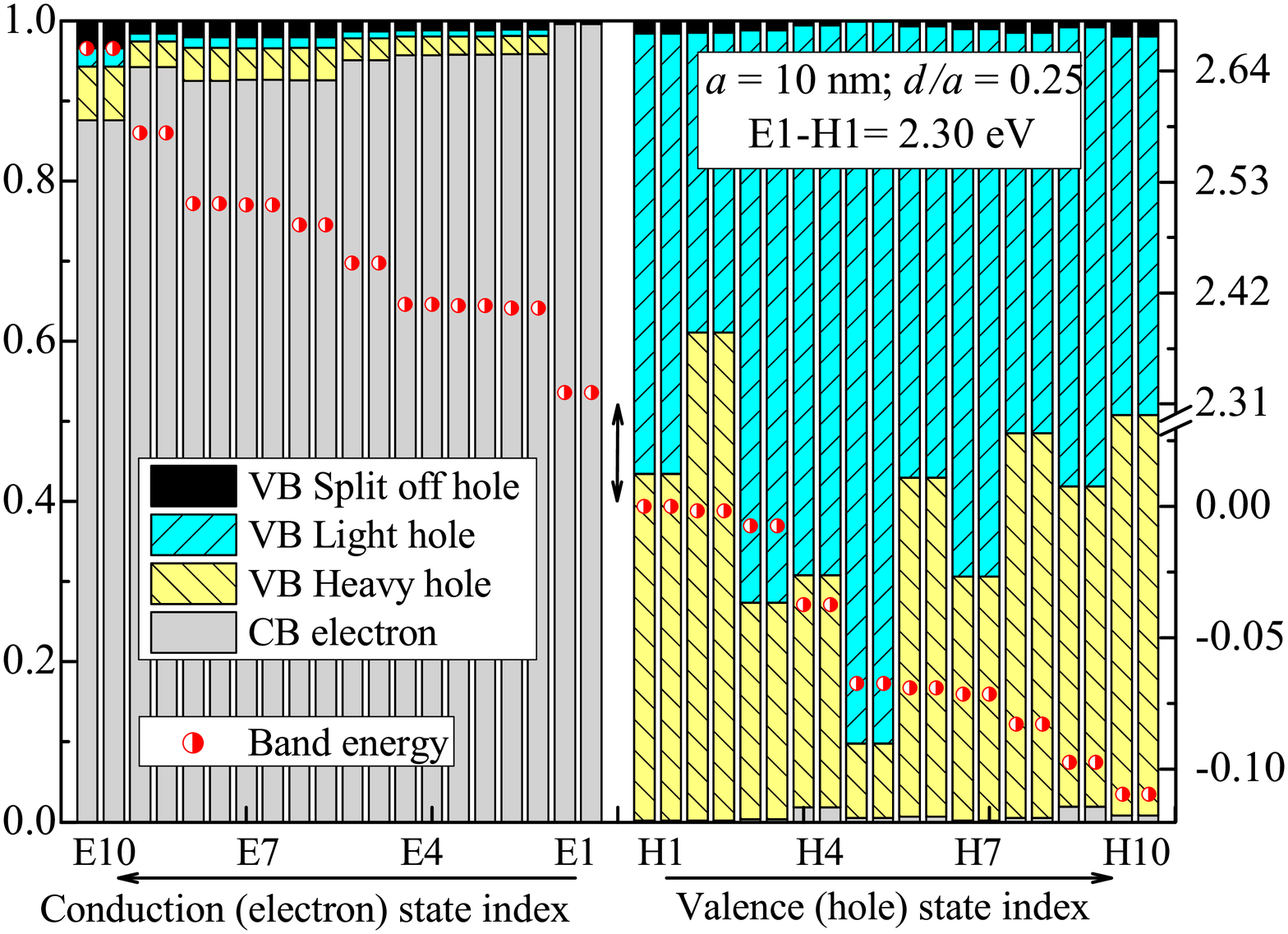}\\
\includegraphics[scale=0.257, angle=270 ]{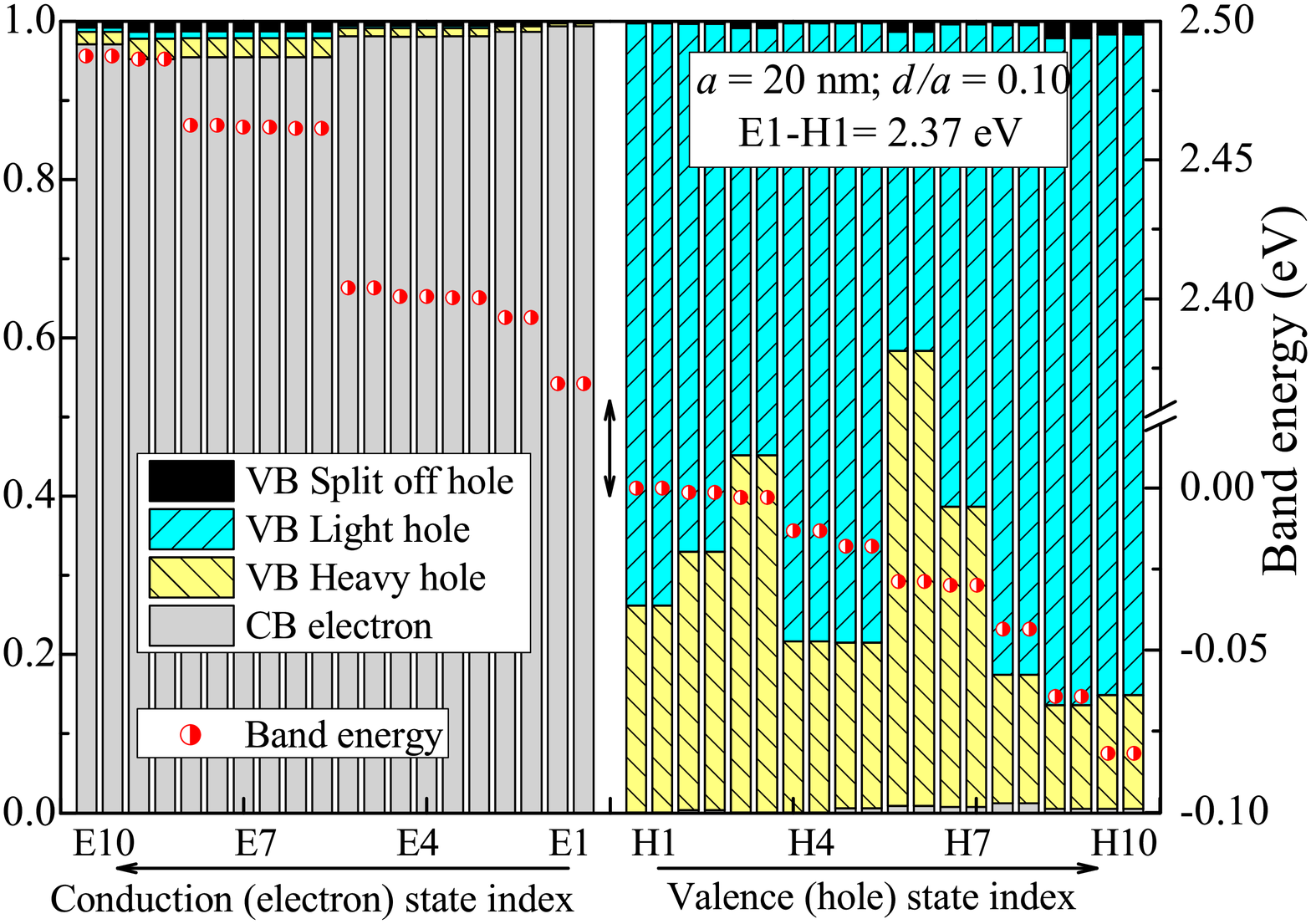}
\includegraphics[scale=0.257, angle=270 ]{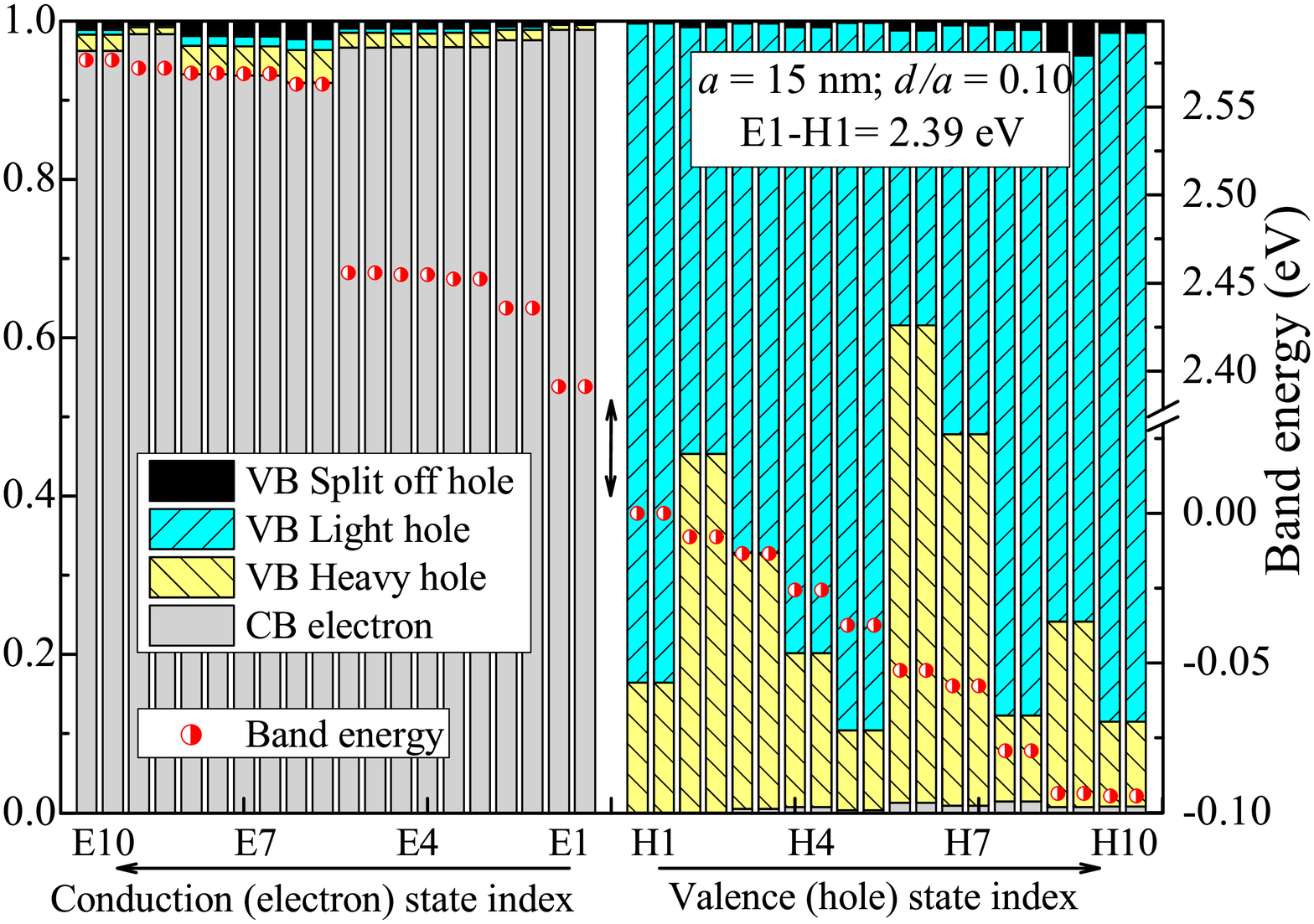}
\includegraphics[scale=0.257, angle=270 ]{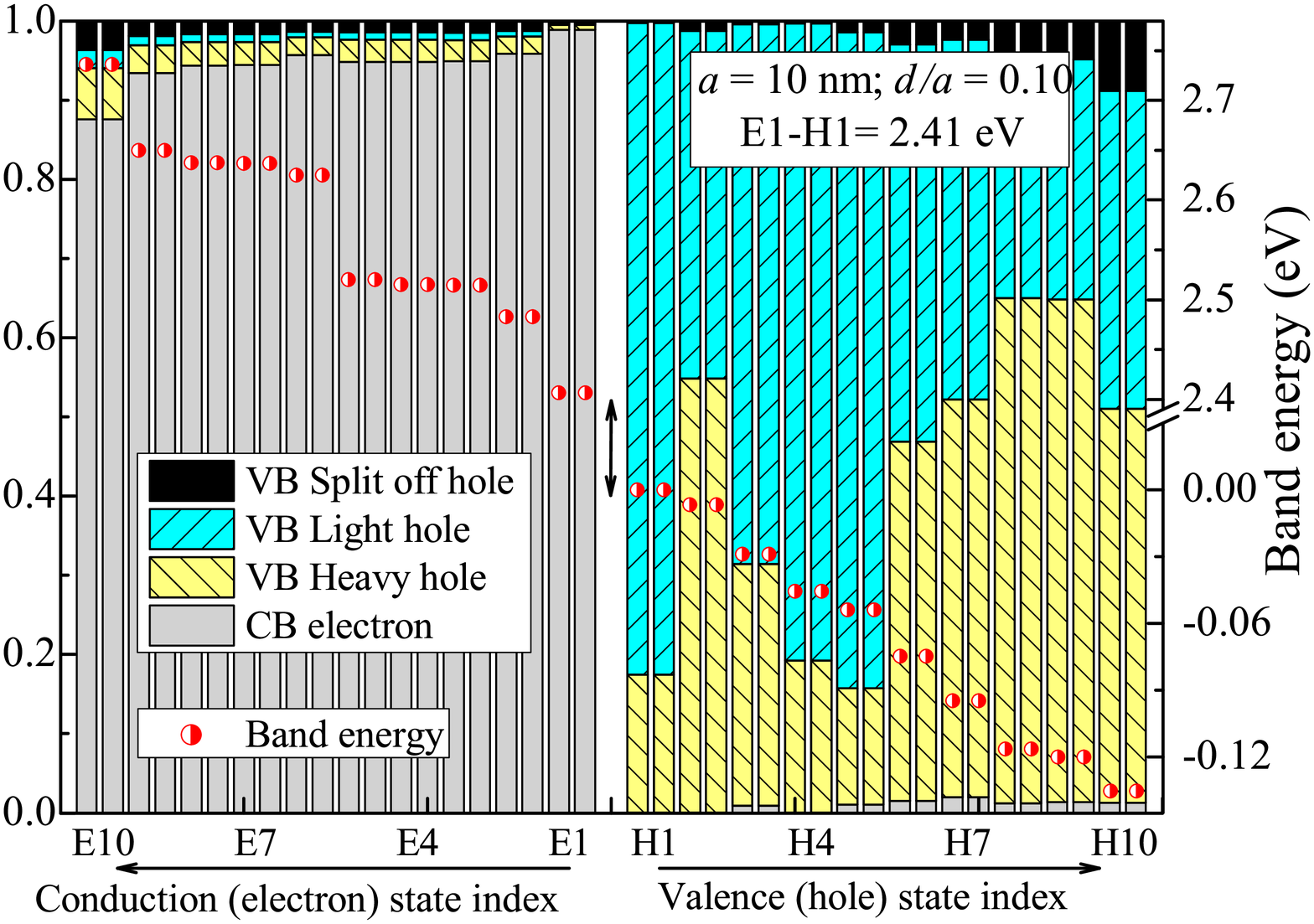}\\

            \caption{Electronic bandstructure and band-mixing probability between CB electrons and VB heavy-, light- and split off holes for 5 ML CdS/CdSe core/crown QRs of lateral size (\textit{a}) = 10, 15 and 20 nm (varying row-wise); and crown width : QR width (\textit{d}/\textit{a}) = 0.40, 0.25 and 0.10 (varying column-wise). Red dots indicate the band energies, while the color bars show the band-mixing probability. Segments with double arrows represent E1-H1 energy gap.}
		\label{fig:EN_Prob_all}
		\end{figure*}

The spatial charge densities shown in Fig.\ \ref{fig:WFS3D-all} are in the \textit{x-y} plane at $z=0$. For most cases, the electron and hole wavefunctions are localized in the CdSe crown, more so for QRs with thinner crowns. As the crown width increases, the wavefunction distribution spreads and leaks into the CdS core gradually. The study of the first ten electron and hole state spatial charge densities is to be done in tandem with the electronic bandstructure and the band-mixing probability between the conduction electrons and valence heavy-, light- and split off-holes due to coupling effect, as shown in Fig.\ \ref{fig:EN_Prob_all}.

For the 15 nm CdS/CdSe QR with (\textit{d}/\textit{a} $=0.10$), both E1 and H1 are \textit{s}-like. E2 is \textit{p}-like, while E3 has \textit{s}-\textit{p}-mixing. E4 and E5 have \textit{p}-\textit{d}-mixing, while E6 and E7 have minute amount of heavy-hole (\textit{hh}) contributions, as inferred from Fig.\ \ref{fig:EN_Prob_all}. Most H states are light-hole (\textit{lh}) dominated, with H2 (53\% \textit{lh}, 45\% \textit{hh}) and H7 (52\% \textit{lh}, 46\% \textit{hh}) having contributions from both \textit{hh} and \textit{lh}, while H6 is \textit{hh} dominated.

The pattern of the first few E states of the 15 nm CdS/CdSe QR with (\textit{d}/\textit{a} $=0.25$) is similar to that with (\textit{d}/\textit{a} $=0.10$) -- E1 being \textit{s}-like, E2 \textit{p}-like, E3 having \textit{s}-\textit{p}-mixing, and so on. Most of the H states are again \textit{lh} dominated, with H2 (48\% \textit{lh}, 50\% \textit{hh}) and H10 (47\% \textit{lh}, 51\% \textit{hh}) having contributions from both \textit{hh} and \textit{lh}, while H8 is \textit{hh} dominated, as can be inferred from Fig.\ \ref{fig:EN_Prob_all}.

Finally, for the 15 nm CdS/CdSe QR with (\textit{d}/\textit{a} $=0.40$), the E1 state is \textit{s}-like, while the E2 has \textit{s}-\textit{p}-mixing. Most of the H states are again \textit{lh} dominated, with H2 (29\% \textit{lh}, 69\% \textit{hh}) and H7 (32\% \textit{lh}, 67\% \textit{hh}) being \textit{hh} dominated, as can be inferred from Fig.\ \ref{fig:EN_Prob_all}. In most cases, the higher E and H states have complex \textit{s}-\textit{p}-\textit{d}-mixing.

In Fig.\ \ref{fig:EN_Prob_all}, we show the electronic bandstructure and the band-mixing probabilities between the conduction electrons and valence
\textit{hh}, \textit{lh}, and split-off (\textit{so}) holes due to coupling effect for varying QR lateral width (\textit{a}) and varying crown width : QR width (\textit{d}/\textit{a}) ratio. The QR lateral widths are varying row-wise between 10, 15 and 20 nm; while the (\textit{d}/\textit{a}) ratios are varying column-wise between  0.40, 0.25 and 0.10. The E1--H1 fundamental transition energy is indicated for each case, as the red dots indicate the band energies. The band-mixing probabilities are indicated using color shaded bars with legends, while segments with double arrows represent E1-H1 energy gap.

    	\begin{figure*}[t]
		\centering
\includegraphics[trim={2cm 0.25cm 2cm 0.25cm},width=0.99\textwidth]{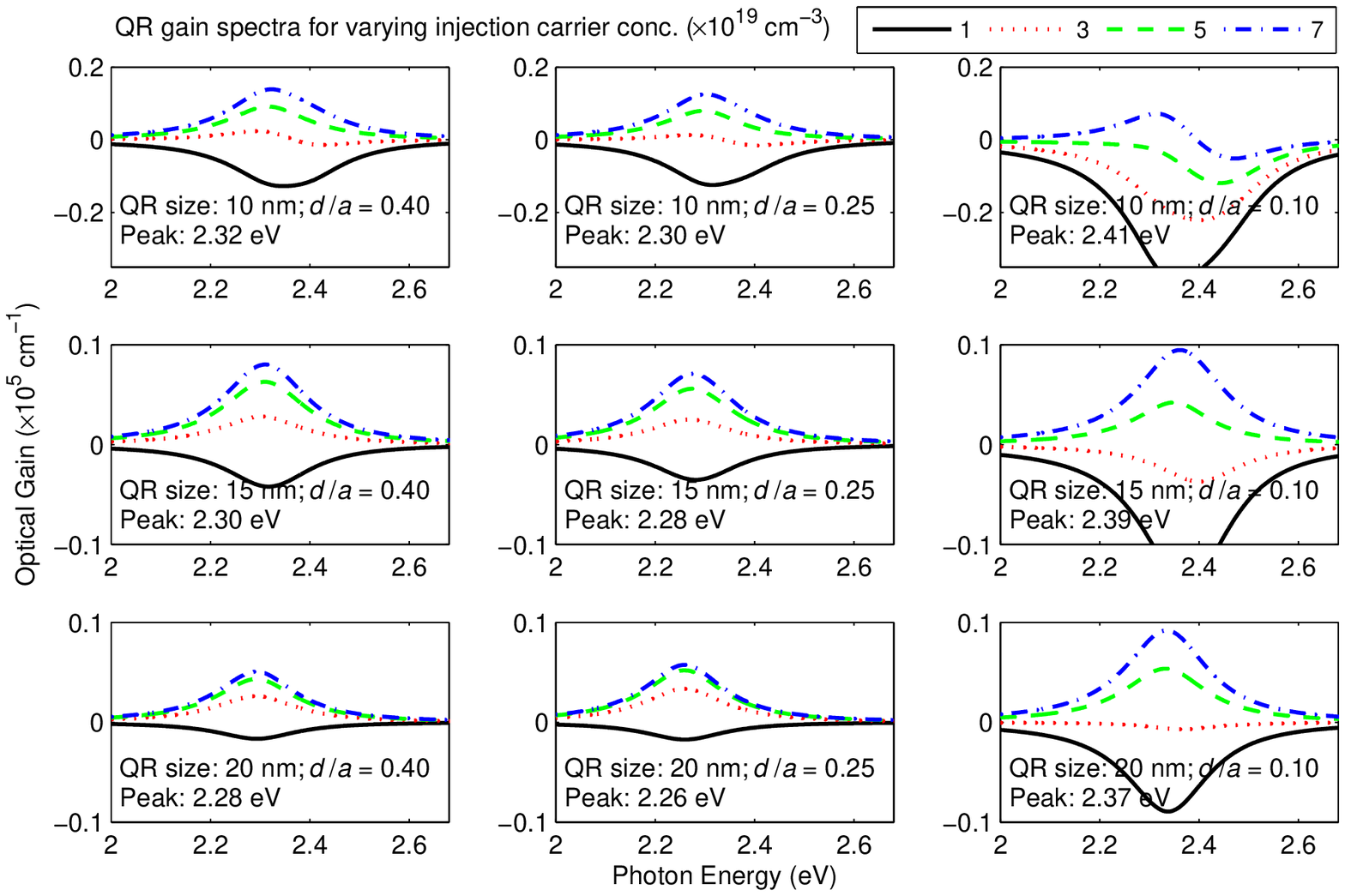}
		\caption{Optical gain spectra (cm$^\text{-1}$) of inverted type-I CdS/CdSe core/crown QRs of lateral size (\textit{a}) = 10, 15 and 20 nm (varying row-wise); and crown width : QR width (\textit{d}/\textit{a}) = 0.40, 0.25 and 0.10 (varying column-wise). For each case, the gain spectra for varying injection carrier concentrations of 1, 3, 5, 7 $\times10^{19}$ cm$^\text{-3}$ is shown as indicated by the legends. Higher carrier concentrations impose a larger gain with a blue-shift in the peak due to band-filling effect.}
		\label{fig:gain_specta_all}
	\end{figure*}
    
The E1--H1 transition energy for each case is in accordance with the values plotted in Fig.\ \ref{fig:EN_TME_plot2} (a). For a fixed (\textit{d}/\textit{a}) ratio, as the size of the QR increases, the transition energy falls because of reduced quantum confinement experienced by the QR. From 10 to 15 nm, and 15 to 20 nm, the reduction is $\sim$200 meV, for all cases. However, for QRs having the same size, as the (\textit{d}/\textit{a}) ratio changes the transition energy pattern is shown and discussed in Fig.\ \ref{fig:EN_TME_plot2} (a). At a broader scale, as the (\textit{d}/\textit{a}) rises, the transition energy decreases, since the fraction of CdSe in the QR system increases while the fraction of CdS decreases -- and CdSe has a lower bandgap compared to CdS. However, at the three (\textit{d}/\textit{a}) values of 0.40, 0.25 and 0.10 we studied, the characteristics has further intricacies. The transition energy pattern is not truly monotonic in the entire range, but has a concavity due to the influence of optical bowing coefficient of CdS$_x$Se$_{1-x}$ = 0.28 eV,\cite{Wei00bowing} [see Fig.\ \ref{fig:EN_TME_plot2} (a)], which induces the minima to occur for an intermediate (\textit{d}/\textit{a}) ratio. The fraction \textit{x} relates to (\textit{d}/\textit{a}) as $x=\left[1-2\left(d/a\right)\right]^2$. We observe that among our three cases, we obtain the least transition energy for (\textit{d}/\textit{a}) = 0.25. As the (\textit{d}/\textit{a}) falls to 0.10, the transition energy rises by $\sim$1100 meV, while as the (\textit{d}/\textit{a}) rises to 0.40, the transition energy rises by only $\sim$200 meV. This follows from the (\textit{d}/\textit{a}) dependent transition energy pattern where the differential transition energy variations are higher for lower values of (\textit{d}/\textit{a}).

Furthermore, we study the span of the first 10 E and H states from Fig.\ \ref{fig:EN_Prob_all}. For any given QR width, as the (\textit{d}/\textit{a}) ratio falls, the span of the first 10 E and H states i.e. $\Delta$E$_{1-10}$ and $\Delta$H$_{1-10}$ spreads out. For instance, considering \textit{a} = 10 nm, $\Delta$E$_{1-10}$ increases from 329 to 334 to 340 meV; and $\Delta$H$_{1-10}$ increases from 63 to 109 to 135 meV, as we move from (\textit{d}/\textit{a}) = 0.40 to 0.25 to 0.10. This is owing to the higher effective hole mass of CdS (0.981) compared to CdSe (0.62); and higher effective electron mass of CdS (0.25) compared to CdSe (0.12), following from the dispersion curve concavity relation $m^*=\hbar^2\left(\frac{\partial^2E}{\partial k^2}\right)^{-1}$. As the (\textit{d}/\textit{a}) ratio falls, the QR is occupied with higher CdS fraction, which induces such an effect. This effect reduces for QRs of larger size.

In Fig.\ \ref{fig:EN_Prob_all}, we have also shown the band-mixing probabilities, that are empirically determined for any QR and depends on the varying extent of coupling between the CB electrons and VB \textit{hh}, \textit{lh} and \textit{so} holes. The reported band-mixing probabilities for each state directly relates to the aggregate electron and hole wavefunction and associated charge densities. For most E states, the primary contributions are from electrons with minor \textit{hh}/\textit{lh}/\textit{so} contributions. But for H states, there are comparable contributions form \textit{hh} and \textit{lh}. For each H state shown in Fig.\ \ref{fig:WFS3D-all}, the single most major contributor as identified from Fig.\ \ref{fig:EN_Prob_all} is mentioned, and for particular cases both \textit{hh} and \textit{lh} are mentioned when their contributions are comparable.

    	\begin{figure*}[t]
		\centering
\includegraphics[width=0.95\textwidth]{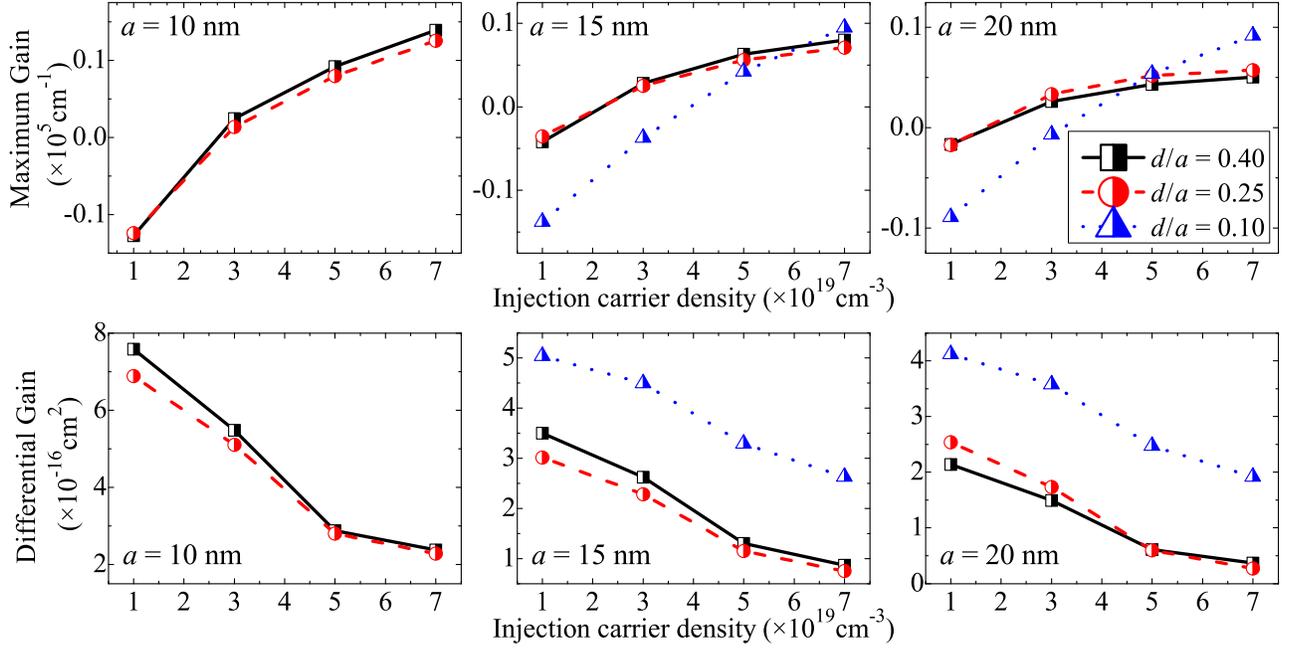}
		\caption{(\textit{top row}): Maximum optical gain, and (\textit{bottom row}): Differential optical gain of inverted type-I CdS/CdSe core/crown QRs as a function of varying injection carrier density of 1, 3, 5, 7 $\times10^{19}$ cm$^\text{-3}$. Each column stands for a fixed value of QR lateral width (\textit{a}) = 10, 15 and 20 nm and each frame shows varying crown width : QR width (\textit{d}/\textit{a}) = 0.40, 0.25 and 0.10. QR width is mentioned in each frame, and legends indicate (\textit{d}/\textit{a}) values.}
		\label{fig:NEW_max_diff_Gain2}
	\end{figure*}
    
\subsection{\label{subsec:opt-gain}Optical Gain and Properties}

In Sec.\ \ref{subsec:elec-Bstr} we examined the strain profiles, transition energy, transition matrix element (TME), charge densities and electronic bandstructure of inverted type-I CdS/CdSe core/crown QRs. In this section we will study the optical properties of the QRs. Fig.\ \ref{fig:gain_specta_all} shows the optical gain spectra of the CdS/CdSe core/crown QRs for varying QR lateral width (\textit{a}) and varying crown width : QR width (\textit{d}/\textit{a}) ratio. The QR lateral widths are varying row-wise between 10, 15 and 20 nm; while the (\textit{d}/\textit{a}) ratios are varying column-wise between  0.40, 0.25 and 0.10. This allows us to study the effect of both QR size and crown width simultaneously. For each QR case we have the gain for varying carrier density ranging from 1 to 7$\times$10$^{19}$ cm$^{-3}$. A higher carrier density translates to a higher lasing threshold current density.

The optical gain spectrum depends on several factors such as the TME (depending on E and H wavefunction overlap), carrier density, Fermi factor, QR dimensions and material, dephasing and scattering rate, etc. Also, the optical gain strongly depends on the exciton binding energy. Based on the method described in Sec.\ \ref{subsec:opt-prop-calc}, the emission peak position for each QR gain spectra calculated and mentioned in Fig.\ \ref{fig:gain_specta_all}, which relates to the E1--H1 transition results from Fig.\ \ref{fig:EN_TME_plot2} (a) and Fig.\ \ref{fig:EN_Prob_all}.

There is a red-shift in the emission wavelength with an increase in the QR size. As the QR volume increases, the extent of quantum confinement reduces and the allowed energy levels come closer. Thus the E1--H1 transitions occur at lower energy. We see a reduction in peak gain energy by $\sim$200 meV as the QR width increases from 10 to 15 nm, and 15 to 20 nm, for all (\textit{d}/\textit{a}) cases. However, for QRs having the same lateral size, the (\textit{d}/\textit{a}) ratio dependent peak gain energy characteristics is shown in Fig.\ \ref{fig:EN_Prob_all}. The optical bowing coefficient induced concavity in the transition energy pattern results in the peak gain energy minima to occur at (\textit{d}/\textit{a}) = 0.25 and is blue-shifted for (\textit{d}/\textit{a}) = 0.10 and 0.40, more so for the former than the latter.

However, for any particular QR case, the optical gain increases with carrier density and there is a marginal blue-shift in the peak gain emission energy. With the injection of carriers, higher electronic states further away from the CB-bottom and VB-top start getting occupied and the subsequent transition typically occurs at a higher, blue-shifted E1--H1 energy value. This phenomenon is known as the band filling effect. This effect is more prominent in volumetrically larger QRs compared to smaller QRs. The allowed energy states due to quantum confinement in larger QRs are much closer. So, there is a higher probability of \textit{near}-E1--H1 transitions to occur as more carriers get injected. In smaller QRs, however, the next higher transition after E1--H1 is much further away, which does not contribute much.

Next we study the effect of carrier density on the peak maximum gain of CdS/CdSe QRs, as shown in Fig.\ \ref{fig:NEW_max_diff_Gain2}. With an increase in the injection carrier density, the maximum gain increases since more carriers get injected, which results in increased radiative recombination. Considering the QR having a width of 15 nm as a case, we see that for (\textit{d}/\textit{a}) = 0.40 and 0.25, the maximum gain increment with injection carrier density has a similar gradient, while for (\textit{d}/\textit{a}) = 0.10, the rise in maximum gain is steeper, and has a comparative rapid increment. For a sufficiently high carrier density, it exceeds the maximum gain of the former cases. The crossover energy is observed to decrease with increasing QR size. For \textit{a} = 10 nm, however, the maximum gains for (\textit{d}/\textit{a}) = 0.10 is not shown, since for our range of carrier densities (1 to 7 $\times10^{19}$ cm$^\text{-3}$), the gain spectra is still mainly in the absorption and transparency phase. Even for 7 $\times10^{19}$ cm$^\text{-3}$, the gain spectra is not truly Lorentzian in shape, so its analysis for maximum gain has been excluded in Fig.\ \ref{fig:NEW_max_diff_Gain2}. But for higher carrier densities it approaches the Lorentzian gain spectra.

For a fixed injection carrier density, the maximum gain depends on the QR dimensions and the (\textit{d}/\textit{a}) ratio. The QR thickness and QR volume inversely affect the optical gain [see Eq.\ \ref{Gspb} and \ref{Gspc}]. For a fixed (\textit{d}/\textit{a}) ratio of 0.40 and 0.25, we observe that the maximum gain falls with a rise in the QR lateral size. The (\textit{d}/\textit{a}) = 0.10 case isn't considered once again since the gain spectra is not truly Lorentzian for \textit{a} = 10 nm. Moreover, for QRs of the same size and same injection carrier density, the (\textit{d}/\textit{a}) ratio affects the maximum gain. As the (\textit{d}/\textit{a}) ratio increases, the maximum gain increases. This is because a higher (\textit{d}/\textit{a}) ratio implies a greater fraction of active CdSe crown in the QR, which has a higher TME [see Fig.\ \ref{fig:EN_TME_plot2} (b)] and higher Fermi factor.

The differential gain is yet another interesting aspect of optical properties of semiconductors. It is the measure of a laser's effectiveness to transform injected carriers to photons. A higher differential gain corresponds to a narrower spectral emission width and a greater modulation speed.\cite{riane06,anson99} Also, a laser's resonant frequency is proportional to the square root of the differential gain with respect to the density of carriers.\cite{tomic03} From Fig.\ \ref{fig:NEW_max_diff_Gain2} we can infer how the differential gain of CdS/CdSe QRs varies with carrier density. It is observed to be higher for QRs with (\textit{d}/\textit{a}) = 0.10 compared to those with (\textit{d}/\textit{a}) = 0.25 or 0.40. But as the carrier density increases, the differential gain decreases for all (\textit{d}/\textit{a}) ratios in a similar fashion. At higher carrier densities, the gain of CdS/CdSe core/crown QRs tends to attain saturation. In general, the differential gain is higher for laterally smaller QRs, and decreases as the QR width increases.

\section{\label{sec:sum-conc}Summary and Conclusion}

In this work, we have studied the optoelectronic properties of type-I core/crown CdS/CdSe quantum rings (QRs) in the zincblende phase -- over contrasting QR lateral size and crown width. Our electronic bandstructure calculations are based on the 8-band \textbf{\textit{k$\cdot$p}} method employing the valence force field model for calculating the strain distributions. The optical gain calculations are based on the density-matrix equation, with excitonic considerations.

In the CdS/CdSe QRs, we observed compressive strain in the CdS core, which increases for thicker CdSe crowns. At the core/crown boundary the strain changes abruptly, the magnitude of which is higher for thinner crowns. Moreover, as the QR size increases the E1--H1 transition energy falls due to reduced quantum confinement. Also, with increasing crown width, we have a greater fraction of CdSe in the QR, and the E1--H1 transition energy falls, but non-monotonically due to the effect of bowing parameter. Furthermore, QRs transformed into pure NPLs have higher TMEs due to localized electron-hole wavefunction overlap in a single material. But for CdS/CdSe QRs, the TME increases with the CdSe crown width where most of the electron-hole wavefunctions are bound, as can be inferred from the spatial charge distribution patterns. For QRs with thicker crowns, the wavefunctions have a greater spatial spread. In the electronic bandstructure, the energy span of the E and H states rises as the crown thickness falls, which is owing to the higher effective electron and hole mass of CdS over CdSe. In all the H states, the major contributing hole-type has been identified.

In the optical gain spectra, the peak emission position experiences a red-shift as the QR width increases owing to reduced confinement. For a given QR width, however, the peak emission position for varying crown widths are in accordance to the respective calculated transition energies. But, for a given QR case, with increasing carrier density, there is slight blue-shift due to band-filling effect, and a rise in the maximum gain and a reduction in the differential gain.

\begin{acknowledgments}
We acknowledge the Ministry of Education, Singapore (RG 182/14 and RG 86/13), A*STAR (1220703063), Economic Development Board (NRF2013SAS-SRP001-019) and Asian Office of Aerospace Research and Development (FA2386-17-1-0039).
\end{acknowledgments}

\bibliography{_qr-paper-refs}

\end{document}